\documentclass[]{aastex631}
\usepackage{natbib}
\usepackage{amsmath}

\newcommand\Teff{T_{\rm eff}}

\newcommand\Msun{\rm M_{\odot}}
\newcommand{\lsun}{L_{\odot}}
\newcommand{\rsun}{R_{\odot}}
\newcommand{\amlt}{\alpha_{\rm MLT}}
\newcommand{\afe}{{\rm[\alpha/Fe]}}

\graphicspath{{./}{figures/}}

\accepted{}
\submitjournal{ApJS}

\shorttitle{MIST2}
\shortauthors{Dotter et al.}

\begin{document}

\title{MESA Isochrones and Stellar Tracks (MIST) II. Models with $\alpha$-enhanced chemical composition}

\author[0000-0002-4442-5700]{Aaron Dotter}
\affiliation{Department of Physics and Astronomy, Dartmouth College, 6127 Wilder Laboratory, Hanover, NH 03755, USA}

\author[0000-0002-4791-6724]{Evan B. Bauer}

\author[0000-0002-8435-9402]{Minjung Park}

\author[0000-0002-1590-8551]{Charlie Conroy}
\affiliation{Center for Astrophysics $|$ Harvard \& Smithsonian, 60 Garden Street, Cambridge, MA 02138, USA}

\author[0000-0001-7506-930X]{Antonino P.\ Milone}
\affiliation{Dipartimento di Fisica e Astronomia “Galileo Galilei”, Universita di Padova, Vicolo dell’Osservatorio 3, Padova 35122, Italy}
\affiliation{Istituto Nazionale di Astrofisica – Osservatorio Astronomico di Padova, Vicolo dell’Osservatorio 5, Padova 35122, Italy}

\author[0000-0002-8717-127X]{Meridith Joyce}
\affiliation{University of Wyoming, 1000 E University Ave, Laramie, WY USA}

\author[0000-0002-8171-8596]{Matteo Cantiello}
\affiliation{Center for Computational Astrophysics, Flatiron Institute, New York, NY 10010, USA}
\affiliation{Department of Astrophysical Sciences, Princeton University, Princeton, NJ 08544, USA}

\begin{abstract}
We update and expand the \texttt{MESA} Isochrones and Stellar Tracks (MIST) database to include variations in the $\alpha$-capture elements, specifically [$\alpha$/Fe]=$-0.2$, 0, +0.2, +0.4, and +0.6 for $-3 \leq$ [Fe/H] $\leq +0.5$.  Variations in [$\alpha$/Fe] are 
included in a self-consistent manner from the stellar interior models to the synthetic spectra used to translate these models in the observational plane.
We describe a number of updates to the physics utilized in these models as well as new information provided by the models.  We validate the models with comparisons
to other stellar evolution models including the previous generation of MIST and other models from the literature.  MIST data products including stellar evolutiont tracks,
isochrones, and bolometric correction tables can be obtained from the MIST project website, \url{https://mist.science}.  All necessary files to reproduce MIST models are available from Zenodo.
\end{abstract}



\section{Introduction} \label{sec:intro}

Stellar models form the backbone of many areas of astrophysics, from the interpretation of the highest redshift galaxies to the determination of exoplanet masses and radii.  These models have been under a constant state of development since computer programs were designed to solve the equations of stellar structure in the early 1950s \citep{1952ApJ...116..317O}.  Modern research efforts involve computing the evolution of large grids of stars from the an initial condition through advanced phases of stellar evolution, including heavy-element nucleosynthesis for massive stars and through thermal pulses during the asymptotic giant branch (AGB) phase for lower-mass stars \citep[e.g.,][]{Marigo2013,Choi16, Marigo2017,Joyce2024}.

To be maximally useful, grids of stellar models are expected to cover a wide range in age, evolutionary phase, initial mass, and initial [Fe/H].  In the case of [Fe/H], it is sometimes assumed that the relative abundance ratios of individual elements is fixed to the adopted solar mixture so that one single parameter, the mass fraction of all metals ($Z$), determines the overall element abundance pattern.  However, it is well known that stars and stellar systems exhibit abundance patterns that deviate substantially from the (scaled) solar mixture.  This is expected because different sites for nucleosynthetic enrichment --- core-collapse supernovae,  thermonuclear detonation of white dwarfs, shell-burning in AGB stars --- produce different abundance patterns \citep{1984psen.book.....C}.  The most common observed deviation from the solar mixture is a change in the $\alpha$-capture elements (O, Ne, Mg, Si, S, Ar, Ca, Ti) with respect to the Fe-peak elements.  This deviation is often parameterized as $\afe$, the ratio of $\alpha$-capture elements compared to Fe elements relative to the solar ratio. Note that the term ``$\alpha$-enhanced'' is used loosely to refer to models within which [$\alpha$/Fe] $\neq 0$.

Stars and stellar systems that display non-scaled-solar abundance patterns are ubiquitous.  In the Milky Way, the majority of stars with [Fe/H]$<-1$ are $\alpha$-enhanced \citep{Tolstoy09}, as are nearly all globular clusters \citep{Pritzl05}.  Populations with non-solar abundance patterns are common among dwarf galaxies in the Local Group, as well as the most massive galaxies in the universe \citep{Worthey1992,Worthey1994,Trager00a,Conroy13b}.There is even evidence that massive star-forming galaxies at $z\sim2$ harbor $\alpha-$enhanced stellar populations \citep[e.g.,][]{Bensby2014, Steidel2016, Cullen2019, Topping2020, Kashino2022}.

These facts motivate the need for large grids of stellar evolution models that span a range in [$\alpha$/Fe] as well as [Fe/H].  Recent grids of $\alpha$-enhanced stellar models have mostly focused on the lower masses and older ages relevant for globular clusters \citep[e.g.,][]{Dotter08b,2014ApJ...794...72V,2018MNRAS.476..496F}. Recent grids covering a wider range in masses and ages include BaSTI \citep{2018ApJ...856..125H,2021ApJ...908..102P} and PaRSEC \citep{Bressan2012,Chen2014,Nguyen2022}.   Few, if any, previous models exist that span the full range in masses (e.g., $0.1-100\Msun$), metallicities including both [Fe/H] and [$\alpha$/Fe], and evolutionary phases.  The main purpose of this paper is to provide such an expansive grid.

A breakthrough in stellar evolution modeling has occurred over the past 15 years thanks to the Modules for Experiments in Stellar Astrophysics (\texttt{MESA}) software instrument \citep{Paxton11, Paxton13, Paxton15, Paxton18, Paxton19, 2023ApJS..265...15J}. In previous work we employed \texttt{MESA} to build a comprehensive grid of stellar evolution tracks and isochrones, referred to as \texttt{MESA} Isochrones and Stellar Tracks (MIST) \citep[][hereafter Papers 0 and 1, respectively]{Dotter16, Choi16}.  These models evolved stars with masses from $0.1-300\Msun$ from pre-main sequence contraction through either central carbon burning or white dwarf cooling, depending on the initial mass, and cover scaled-solar compositions from [Fe/H]$=-3$ to $+0.5$.  

In this work we update the MIST database in two important respects. First, we describe updates to the stellar physics. Second, we expand the coverage beyond solar-scaled compositions to include a range of $\alpha$-enhancements from [$\alpha$/Fe]=$-0.2$ to $+0.6$ in steps of 0.2 dex.  These new models are available for download from the MIST project webpage\footnote{\url{http://mist.science}} and Zenodo.


\section{The MIST Stellar Models} \label{sec:models}
Paper 1 gives a detailed description of the adopted physics, notably including mixing processes, rotation, and mass loss, used in the \texttt{MESA} models that form the foundation of MIST, as well as information about how those tracks are converted into isochrones (see also Paper 0), and how both types of models are transformed into the observational plane.

In this paper, the algorithmic treatment of microphysics (equation of state [EOS] and opacities), nucleosynthesis, mixing, rotation, and mass loss are essentially the same as described in Paper 1, except where we describe extensions of the microphysics in the subsections below. We therefore do not exhaustively revisit all microphysics inputs in detail; the reader is referred to Paper 1 for topics not covered here, as well as to the repository of codes shared with this paper that enable full reproduction of all stellar modeling (see Section~\ref{sec:MESA}).

\subsection{Abundances} \label{sec:abund}
We use the term $\afe$ to be representative of the $\alpha$-capture elements. It is not possible to express a single definition of ``$\alpha$-enhancement'' that works for all stars in all environments. Instead, stellar models of $\alpha$-enhanced stars have chosen a number of different patterns.  In the past, a variety of different $\alpha$-enhancement patterns were used to compute stellar evolution models, notably the mixture adopted by \citet{SalarisWeiss1998} that was used in previous generations of the Padova \citep{Salasnich2000} and BaSTI models \citep{Pietrinferni2006}. However, the latest grids from Victoria-Regina \citet{2014ApJ...794...72V}, BaSTI \citep{2021ApJ...908..102P}, and Dartmouth \citep{Dotter08b} adopt an approach where each $\alpha$-element is enhanced by the same amount relative to its scaled-solar abundance, and that value is equal to [$\alpha$/Fe]. This fixed approach is adopted in the present work as well.  

It is worthwhile to note that while the $\alpha$-enhancement pattern used by Dartmouth, Victoria-Regina, and BaSTI is the same as in this work, the adopted solar abundance patterns differ: Dartmouth used \citet[][hereafter GS98]{Grevesse98}, BaSTI used \citet{Caffau11}, and Victoria-Regina adopted the \citet{Asplund09} solar mixture.  While Paper 1 adopted the \citet{Asplund09} solar abundances, we have instead adopted \citet{Grevesse98} in this paper for two reasons: first, the latest analyses of the solar abundances are tending back towards solar metal abundances consistent with GS98 \citep{2022A&A...661A.140M,2023A&A...672L...6P} and, second, there are high- and low-temperature opacities readily available for the GS98 composition with $\alpha$-element enhancements.

The MIST model grids include 15 [Fe/H] values spanning $-3 \leq $ [Fe/H] $\leq +0.5$ and 5 $\afe$ values spanning $-0.2 \leq \afe \leq +0.6$ for a total of 74 unique chemical compositions. The grid spacing is 0.2 dex in $\afe$ and 0.25 dex in [Fe/H], except for [Fe/H] $<-3$ where the spacing is 0.5. We exclude the grid at [Fe/H] $=+0.5$, [$\alpha$/Fe] $=+0.6$ due to the lack of necessary input physics.

In the following, we relate the mass fraction of helium ($Y$) and all heavier elements ($Z$) used in our models. The helium  enrichment with respect to heavier elements, $\Delta$Y/$\Delta$Z, expressed as
\begin{equation}
    \frac{\Delta\,Y}{\Delta\,Z}=\frac{Y_{\odot,{\rm init}}-Y_p}{Z_{\odot,{\rm init}}-0} = \frac{0.0245}{0.0185} \approx 1.32
\end{equation}
where the subscript $p$ refers to the primordial value derived from Big Bang Nucleosynthesis \citep[$Y_p=0.249$,][]{Planck15} and the $\odot{\rm , init}$ symbol refers to the initial value of our solar-calibrated model (see Table \ref{tab:solar_calib}). 
 This value is lower than that used in Paper 1 ($\Delta$Y/$\Delta$Z=1.5) because of the chosen proto-solar abundances (see Section 1 and Table 4 of Paper 1).


\subsection{MESA} \label{sec:MESA}

For the present work we use the \texttt{MESA} software instrument \citep{Paxton11,Paxton13,Paxton15,Paxton19,2023ApJS..265...15J} version 11701 \citep{MESA_11701} supported by the \texttt{MESA} SDK \citep{MESASDK}.  We use radiative opacities from OPAL \citet{Iglesias1996} and \citet{Ferguson2005}; nuclear reaction rates from JINA REACLIB \citep{REACLIB}; the EOS is a combination of OPAL \citep{Rogers2002}, SCvH \citep{Saumon1995}, HELM \citep{Timmes2000}, and PC \citep{Potekhin2010} as detailed in \citet{Paxton11,Paxton19}.
We have utilized EOS tables computed with the FreeEOS\footnote{https://freeeos.sourceforge.net/documentation.html} software instrument to supplement the partially-ionized parameter space otherwise covered by the OPAL EOS for metal mass fraction $Z \geq 0.04$.  The FreeEOS tables were provided for later versions of \texttt{MESA} as described by \citet{2023ApJS..265...15J} and backported into \texttt{MESA} version 11701.  The version of \texttt{MESA} used in this paper was modified to fix a couple of issues that were discovered in newer versions of \texttt{MESA} as well as support for the aforementioned FreeEOS tables, atmosphere boundary condition tables, and extended sets of Rosseland mean opacity tables to properly reflect the $\afe$ variations covered by the MIST model grid. (All references to opacity in this paper refer to the Rosseland mean opacity.) In order to enable reproducibility of all our stellar modeling, we make available our entire \texttt{MESA} installation including modifications \citep{MIST_11701} as well as work directories, inlists, and all other input files \citep{MESA_work}.
An elucidation of the changes made is given in the Appendix.

The impact of $\afe$ variations on stellar structure can enter our models through multiple aspects of the microphysics, including the opacities, EOS, atmosphere boundary conditions, and nuclear reactions. The treatment of nuclear reaction networks in \texttt{MESA} accounts for the full composition vector of the model, so nuclear reactions will naturally incorporate the impact of $\afe$ variations, but the treatments for other aspects of microphysics mentioned above all contain components that rely on interpolating tables that by default assume solar-scaled metallicity mixtures. We therefore supplement the default treatments for tabulated opacities and atmosphere boundary conditions with additional tabulations to capture $\afe$ variations. For opacities, we used additional OPAL opacity tables for each of the 5 $\afe$ values covered by our model grids, using the GS98 abundance mixture with $\alpha$-capture elements scaled as described in Section~\ref{sec:abund}. For atmosphere boundary condition tables, we describe our supplemental treatment in Section~\ref{sec:atm}. The low-temperature opacities and the atmosphere boundary conditions are responsible for mediating by far the largest impact of composition variations on our stellar interior models \citep{2006astro.ph..5666W, Dotter2007, VandenBerg12}.

\subsection{The Atmosphere Boundary Condition} \label{sec:atm} 
As in Paper 1, we use a hybrid approach to setting the atmosphere boundary condition in our stellar evolution models. For stellar models with ${\rm M \geq 8 M_{\odot}}$, we use the \texttt{simple\_photosphere} option in \texttt{MESA}---this amounts to the (simple) assumption that the opacity in the atmosphere is a constant that can be pulled out of the  T-$\tau$ integration.

For stellar models with ${\rm M < 8 M_{\odot}}$, we use the T-$\tau$ integration method described by \citet{2022RNAAS...6..112D}. The T-$\tau$ relation refers to a relation between temperature (T) and Rosseland mean optical depth ($\tau$) in the stellar atmosphere. The T-$\tau$ relation from \citet{Vernazza1981} is adopted.  The method extends the atmosphere boundary condition T-$\tau$ integration by checking the local conditions for convection at each step in the numerical integration and substituting the values derived from the mixing length theory (MLT) when convection is present. This method enables the matching point between the atmosphere integration and the interior model to be safely made at a point where the model is convective. As pointed out by \citet{Chabrier97} there is considerable advantage to choosing the match point between atmosphere and interior within the surface convection zone (if there is one). The fitting point between atmosphere integration and interior model is chosen to be $\tau=10$, which is confirmed to be within the surface convection zone for all models that have a convective surface layer.  The T-$\tau$ integrations are tabulated according to surface $\Teff$ and $\log(g)$, analogous to the way that model atmosphere boundary condition tables are structured.  Individual tables are made for each of the distinct chemical compositions in this paper. These tables are used within \texttt{MESA} in place of the default set of tabulated atmosphere boundary conditions.

\subsection{Rotation}
As in Paper~1, we include two sets of models: one with and one without rotation. Models with rotation are initialized with angular velocity $\Omega = 0.4 \Omega_{\rm crit}$ near the Zero Age Main Sequence (ZAMS) for $M_{\rm init} \geq 1.8 \Msun$, with rotation ramped\footnote{The `ramp' function mentioned here and in the following paragraph is sinusoidal, rather than linear, in nature.  The purpose of this is to affect the ramp without introducing a discontinuity in the first derivative.} down to zero for models with $M_{\rm init} \leq 1.2 \Msun$. Here $\Omega_{\rm crit}$ is the critical (Keplerian) angular velocity.
Modeling of rotation includes both angular momentum transport and associated compositional mixing with a relative efficiency factor of $f_c = 1/30$ for compositional diffusivity relative to angular momentum diffusivity \citep[][section 6]{Paxton13}. The rotational mixing processing accounted for include dynamical shear instability, secular shear instability, Solberg-H{\o}iland (SH) instability, Eddington-Sweet (ES) circulation, and Goldreich-Schubert-Fricke (GSF) instability \citep{Pinsonneault1989,Heger2000,Paxton13}.

In addition to these rotational processes that were included in Paper 1, the model grids now also include angular momentum transport due to the Tayler-Spruit mixing \citep{Spruit2002} following the prescription of \cite{Fuller2019}. This additional angular momentum transport allows compact cores to spin down on the red giant branch (RGB) and asymptotic giant branch (AGB) and achieves reasonable agreement with seismologically-inferred spin rates of cores on the RGB and red clump \citep{Mosser2012,Deheuvels2015,Gehan2018}. Because this prescription is primarily calibrated for intermediate mass stars and their white dwarf descendants, we only apply this \cite{Fuller2019} prescription for models with $M < 7\, M_\odot$, with the mechanism fully on for $M<4\,M_\odot$ and ramped down to zero for $4\,M_\odot < M < 7\,M_\odot$. By the time these models reach the end of the AGB, their cores have slowed sufficiently to evolve into white dwarfs that will generally agree with observed $\sim$day spin periods for white dwarfs \citep{Hermes2017,Fuller2019}.

\begin{figure*}
    \centering
    \includegraphics{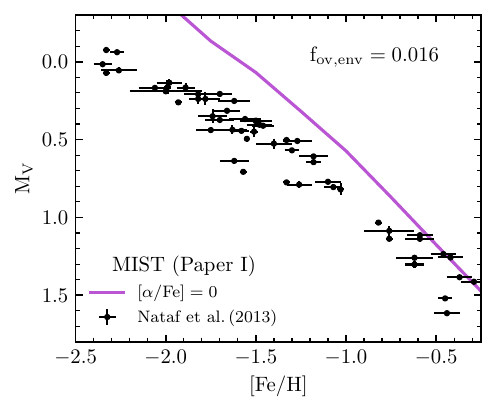}
    \includegraphics{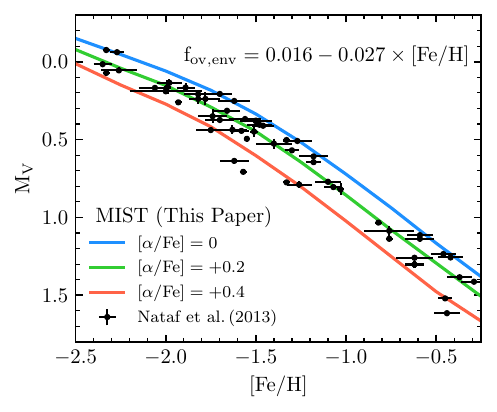}
    \caption{The Red Giant Branch Bump (RGBB) data sample compiled by \citet{Nataf2013} is compared with Paper 1 (left) and this paper (right) at 12 Gyr. Paper 1 models have only scaled-solar composition. The main difference between the two panels is the calibration of $f_{\rm ov,env}$ (see Section $\ref{sec:mixing}$ for details).}
    \label{fig:RGBB}
\end{figure*}

\subsection{Mixing Processes} \label{sec:mixing}
We use the exponential-decay formalism \citep[][equation 2]{Herwig2000}, with adjustable parameter $f$, to model mixing across radiative-convective boundaries.  Overshoot mixing was calibrated in Paper 1, adopting the same parameter $f_{\rm ov}$ for all mixing boundaries. It should be noted that the calibration of overshoot parameters performed in Paper 1 was done using the solar calibration.  The need to perform an additional calibration was motivated by the inability of MIST models to correctly predict the observed trend in the brightness of the red giant branch bump (RGBB), the overdensity of stars created when the H-burning shell briefly meets the bottom of the surface convection zone during red giant evolution, with metallicity (see Figure \ref{fig:RGBB}).

Using the data set compiled by \cite{Nataf2013}, we have undertaken to calibrate the overshoot parameter specifically for the case of mixing below the surface convection zone, $f_{\rm ov,env}$. The procedure involved computing sets of stellar models with our standard configuration while varying only $f_{\rm ov,env}$ at a range of metallicities corresponding to the \citet{Nataf2013} dataset. With these sets of models computed, it was possible to a relation between $f_{\rm ov,env}$ and [Fe/H] that follows the trend seen in the data, 
\begin{equation}
f_{\rm ov,env} = 0.016 - 0.027 \times {\rm [Fe/H]}
\end{equation} 
with the further constraint that $0.01 \leq f_{\rm ov,env} \leq 0.08$ because of the limited [Fe/H] range of the \citet{Nataf2013} sample compared to the [Fe/H] range of the models.  The value of $f_{\rm ov,env} = 0.016$ is the solar-calibrated value from Paper 1.

Figure \ref{fig:RGBB} demonstrates the result of the $f_{\rm ov,env}$ calibration.  The left panel shows the trend with [Fe/H] for Paper 1 models and the right panel shows the trends from this paper, including a range of $\alpha$-enhancement.  The result of the calibration described here is that the models follow the observed trend over the full range of [Fe/H], whereas the uncalibrated models from Paper 1 predicted a much steeper trend, with the bump magnitude being over-predicted by roughly 0.5 for the most metal-poor globular clusters.

Paper 1 implemented the `radiation turbulence' model \citep{Morel2002} to inhibit atomic diffusion near the stellar surface.  In this paper the surface mixing has been changed to the turbulent diffusion treatment described by \citet{VandenBerg12}; for discussion see Section 2.1 of \citet{Dotter2017}.

\subsection{Solar Calibration}\label{sec:solar}
The changes with respect to Paper 1 summarized above, especially the change of solar abundance pattern, necessitated a revised solar calibration. We adopt the nominal solar values of IAU Resolution 2015.B3 \citep{IAU.2015.B3}. The results of the solar calibration and the initial parameters of the solar model are summarized in Table \ref{tab:solar_calib}. 

\begin{table}
\centering
\begin{longtable}{l l l} \\ 
\caption{Solar calibration results for MIST2} \\
\hline
\hline \noalign{\smallskip}
Parameter & Target & Model Value \\ \noalign{\smallskip}
\hline
\noalign{\smallskip}
$\log(L/\lsun)$ & 0 & $3.72\times10^{-6}$ \\
$\log(R/\rsun)$ & 0 & $-5.33\times10^{-5}$ \\
 $Y_{\rm surf}$ & $0.2485$\footnote{\label{B04}\citet{Basu2004}} & 0.2482 \\
 ${\rm [Fe/H]}_{\rm surf}$ & 0\footnote{GS98} & 0.003 \\ 
 $R_{\rm cz}/R_{\odot}$ & 0.7133\footnote{\citet{Basu2004}} & 0.7174  \\ \noalign{\smallskip}
\hline
\noalign{\smallskip}
$\amlt$ & \nodata & 2.03  \\
$f_{\rm ov,\;env}$ & \nodata &  0.016 \\
$X_{\rm initial}$ & \nodata & 0.7080  \\
$Y_{\rm initial}$ & \nodata & 0.2735  \\
$Z_{\rm initial}$ & \nodata & 0.0185  \\
\noalign{\smallskip}
\hline \hline
   \label{tab:solar_calib}
\end{longtable}
\tablecomments{The top half compares observed and modeled values at the solar age.  The bottom half lists solar-calibrated model parameters. $X$, $Y$, and $Z$ refer to the mass fractions of hydrogen, helium, and all heavier elements, respectively. Subscript `surf' refers to the surface value of the model. Subscript `cz' refers to the location of the bottom of the surface convection zone. $\alpha_{MLT}$ is the mixing length parameter and $f_{\rm ov,env}$ is described section 2.5.}
\end{table}

\subsection{Evolutionary Coverage}
Paper 1 described evolutionary models that covered from the pre-main sequence to either the end of the main sequence (for very-low-mass stars), to the white dwarf cooling sequence (for low- and intermediate-mass stars), or to the end of central C-burning (for high-mass stars).  In this paper evolutionary coverage is the same for very-low-mass and high-mass stars.  However, for low- and intermediate-mass stars we have stopped the evolution once the models reach the end of the thermally-pulsating asymptotic giant branch (TP-AGB) phase, having shed the majority of the H-rich envelope and begun to evolve to higher $\Teff$.  

Significant improvements to the treatment of white dwarf models in \texttt{MESA} versions subsequent to \texttt{MESA} 11701 motivate a separate study dedicated to extending the current evolutionary tracks onto the white dwarf cooling curve \citep{2023ApJ...950..115B,2023ApJS..265...15J}. The low and intermediate mass models therefore end at the post-AGB phase with their core composition largely set, ready to evolve over to and down the cooling sequence as white dwarfs.  Updated white dwarf cooling models in the MIST framework is described in a companion paper (Bauer et al., accepted to ApJ).

\subsection{Evolutionary Tracks and Isochrones}
Evolutionary tracks and isochrones are constructed using the `equivalent evolutionary phase' (EEP) formalism and the \texttt{iso} code \citep{2016ascl.soft01021D}, described in Paper 0. Stellar evolution tracks computed by \texttt{MESA} are converted into EEP-based tracks and then those EEP-based tracks are interpolated into isochrones.  Tracks and isochrones provided on the MIST project webpage use the same file format as in Paper 1.  Some additional data columns have been added, as described in Section~\ref{sec:mixing}.  The file format is otherwise unchanged from Paper 1.

\subsection{Synthetic Spectra and Bolometric Corrections}\label{sec:spectra}
Our stellar evolution tracks and isochrones specify physical properties ($\Teff$, $\log(g)$, composition) at the surface, which must then be transformed into observables through synethetic spectra and bolometric corrections for a variety of filter sets. We accomplish this with a custom grid of stellar spectral models calculated using the 1D LTE plane-parallel atmosphere and radiative transfer codes {\tt ATLAS12} and {\tt SYNTHE} maintained by R.~Kurucz \citep{Kurucz1970,Kurucz1981,Kurucz1993, Kurucz11}. The line list used in the radiative transfer calculation was provided by R.~Kurucz and has been empirically tuned to the observed, ultra-high-resolution spectra of the Sun and Arcturus, including the ExoMol TiO database from \citet{2019MNRAS.488.2836M}. We adopt a constant microturbulence of $v_{\rm micro} = 1 \, \rm km\, s^{-1}$. The grid is computed with $\Delta \log g = 0.5$ over the range $[-1,5]$, $\Delta [{\rm Fe/H}] = 0.25$ over the range [$-3,+0.5$], $\Delta \afe = 0.2$ over the range [$-0.2$,+0.6], and an irregular (approximately log-spaced) $\Teff$ grid with 40 points between 3500 and 15,000~K. We compute synthetic spectra at a resolving power of $R = 1{,}000{,}000$, and then subsequently smooth to a resolution of $R = 42{,}000$.
The abundance patterns used in the stellar atmospheres and interior models are identical, as much as possible, considering that the model atmospheres and synthetic spectra include abundances of many more elements than are directly modeled by the stellar interior code.

We adopt the IAU resolution 2015.B2 \citep{IAU.2015.B2} recommended value for the zeropoint of the bolometric magnitude scale, in which the adopted nominal solar luminosity corresponds to an absolute bolometric magnitude
for the sun of $M_{\rm Bol,\odot}, = 4.74$. Bolometric corrections are constructed following \citet{Girardi2008};
the integrals are evaluated in ‘photon counting‘ mode, not energy counting \citep[][]{Bessell2012}. Zeropoints
defining the photometric standard system are chosen individually for each filter set from one of either Vega \citep{Bohlin16}, AB \citep{1974ApJS...27...21O}, or ST \citep{1986HiA.....7..833K}. In all regards this approach is identical to Paper 1. Extinction is incorporated inside the integral over wavelength for each filter and each synthetic spectrum. For this paper we have included the extinction
curve derived by \citep{2007ApJ...663..320F} with selective extinction $R_V=3.1$ as well as a range of extinction $A_V$ from 0 to 6. An up-to-date list of supported photometric systems is provided on the MIST webpage.

\subsection{New features}
A few new data columns have been added to the stellar evolution tracks and isochrones.  A README file including descriptions of all table columns is provided on the MIST webpage.  The following is a description of new columns added since Paper 1.

\subsubsection{Gravity darkening}
\citet{Paxton19} introduced the capability to include the effects of gravity darkening in \texttt{MESA} rotating stellar evolution models via the theoretical framework of \citet{2011A&A...533A..43E}.  MIST stellar evolution tracks and isochrones have been expanded to include the luminosity and effective temperature that would be observed if the star were observed pole-on and equator-on in rotating models\footnote{These new columns are present in all data files but are not distinct for non-rotating models.}.  This information can be useful, for example, in estimating the envelope of possible observables for a given rotating stellar model depending on the orientation of the star's spin axis with respect to the observer.

\subsubsection{Surface convection zone}
For stars with a surface convection zone, the tracks and isochrones include both the mass and radius coordinate of both the top and the bottom of zone.  In addition, we provide estimates of the convective turnover time of the surface convection zone.  One estimate is a global average, calculated as $\int \frac{dr}{v_{\rm conv}}$ with $r$ the radial coordinate and $v_{\rm conv}$ the convective velocity; the integral is over the radial extent of the surface convection zone.  The other two estimates are local, and are given by the ratio of the mixing length ($\Lambda=\alpha_{\rm MLT} H_P$) to the convective velocity evaluated locally at one half and one mixing length above the bottom of the surface convection zone.

\subsubsection{Internal structure}
A measure of central concentration, the apsidal motion constant $k_2$ \citep[][]{Kopal1978} can also be measured directly from binary systems and, thus, constitutes a useful probe of stellar models.


\section{Behavior of the models}\label{sec:behave}

In this section, we present an overview of the isochrones on Hertzsprung-Russell (HR) diagrams and color-magnitude diagrams (CMDs), with an emphasis on the impact of the additional $\afe$ dimension of the model grids compared to Paper~I.


\subsection{New isochrones compared to Paper 1}

\begin{figure*}
\centering
    \includegraphics{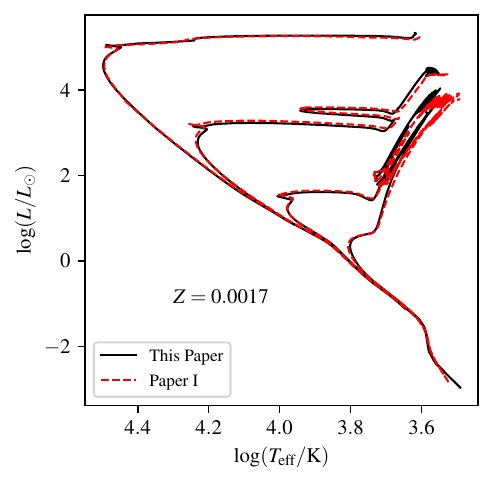}
    \includegraphics{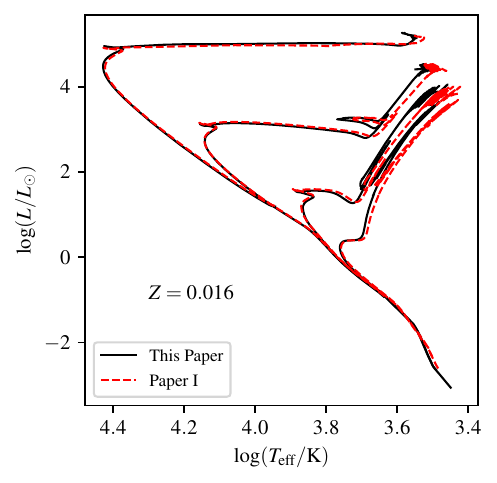}
    \caption{MIST isochrones at 10 Myr (upper left), 100 Myr, 1 Gyr, and 10 Gyr (lower right). The left panel shows $Z=0.0017$ ([Fe/H] $\approx-1$) and the right panel shows $Z=0.016$ ([Fe/H] $\approx 0$), for both Paper 1 (red) and this paper (black). In this and subsequent figures, the pre-MS evolution has been omitted from the younger ages for clarity.}
    \label{fig:MIST2v1_HRD}
\end{figure*}

Figure~\ref{fig:MIST2v1_HRD} shows example isochrones at 10~Myr, 100~Myr, 1~Gyr, and 10~Gyr for two different values of [Fe/H], at fixed $\afe = 0$ for comparison to the solar-scaled isochrones from Paper 1. The only significant difference between versions that manifests in the H-R diagram in Figure \ref{fig:MIST2v1_HRD} is the change of surface boundary condition and solar-calibrated $\alpha_{\rm MLT}$ that influences the temperature scale of the RGB.

\subsection{Sensitivity of the models to [$\alpha$/Fe]}\label{sec:sensitivity}

\begin{figure*}
    \centering
    \includegraphics{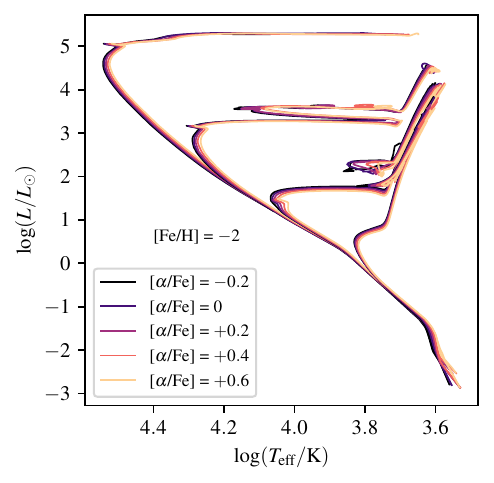}
    \includegraphics{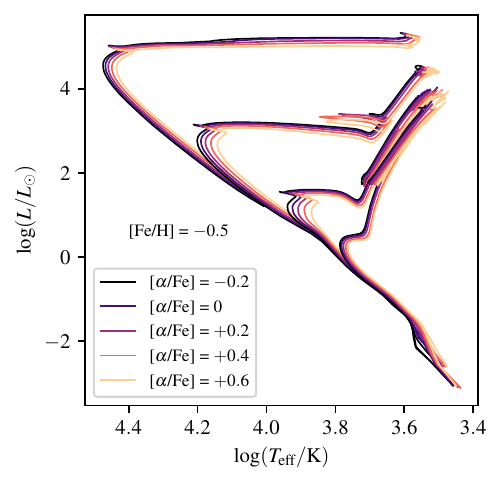}
    \caption{MIST isochrones at 10 Myr, 100 Myr, 1 Gyr, and 10 Gyr, showing the effect of $\afe$ enhancements across the range $-0.2$ to $+0.6$ in the theoretical HR diagram. The left panel show $[{\rm Fe/H}] = -2.0$ and the right panel shows $[{\rm Fe/H}] = -0.5$ isochrones.}
    \label{fig:alpha_HRD_all}
\end{figure*}

As mentioned in Sections~\ref{sec:abund}--\ref{sec:atm} and~\ref{sec:spectra}, the variations in the $\afe$ dimension of composition impact both the physics of stellar evolution tracks and the bolometric corrections. Figure~\ref{fig:alpha_HRD_all} shows two sets of isochrones at fixed [Fe/H] displaying the full range of [$\alpha$/Fe] in the H-R diagram. The influence of [$\alpha$/Fe] is smaller at $[{\rm Fe/H}] = -2.0$ than at $[{\rm Fe/H}] = -0.5$. The overall effect is a combination of factors, including increasing (primarily) the oxygen abundance on the stellar lifetimes and modifying the opacity in the surface layers \citep[see, e.g.,][]{Dotter07}.

\begin{figure*}
    \centering
    \includegraphics{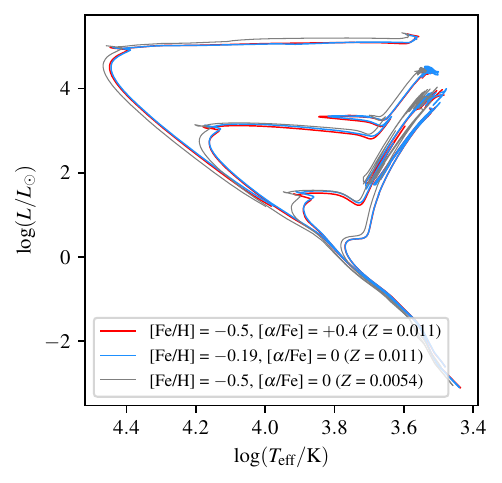}
    \includegraphics{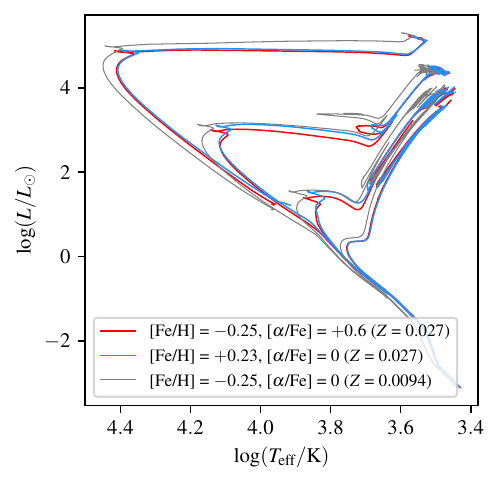}    
    \caption{MIST isochrones at 10 Myr, 100 Myr, 1 Gyr, and 10 Gyr.  The red line is the $\alpha$-enhanced model with [Fe/H]=$-0.5$ (left) or $-0.25$ (right) while the blue line is made from a scaled-solar composition but interpolated to the same total Z as the red curve in each panel. For comparison, the gray lines are scaled-solar models with the same [Fe/H] as the red lines.}
    \label{fig:ZinterpHRD}
\end{figure*}

The possibility of reproducing the influence of $\alpha$-enhancement on stellar evolution tracks and isochrones by (re)scaling the total $Z$ of models with scaled-solar abundances has been discussed in the literature for years \citep{Salaris93}.  As already hinted at in Figure \ref{fig:MIST2v1_HRD}, the challenge becomes greater as total $Z$ increases.  Figure~\ref{fig:ZinterpHRD} shows two examples that demonstrate the value of having $\afe$ and [Fe/H] as independent dimensions. In each panel of Figure \ref{fig:ZinterpHRD}, $\alpha$-enhanced isochrones are compared to two other isochrones at $\afe = 0$, representing solar-scaled mixtures. One of these isochrones has the same value of $[{\rm Fe/H}]$ while the other matches the total $Z$ of the $\alpha$-enhanced isochrone. While the case of [Fe/H]=$-0.5$ shown on the left panel of Figure \ref{fig:ZinterpHRD} dictates that scaling total $Z$ does an adequate job, the case shown on the left shows that the scaled-solar isochrones are not able to capture the morphologies of $\alpha$-enhanced isochrones at older ages.

\begin{figure*}
    \centering
    \includegraphics[width=\textwidth]{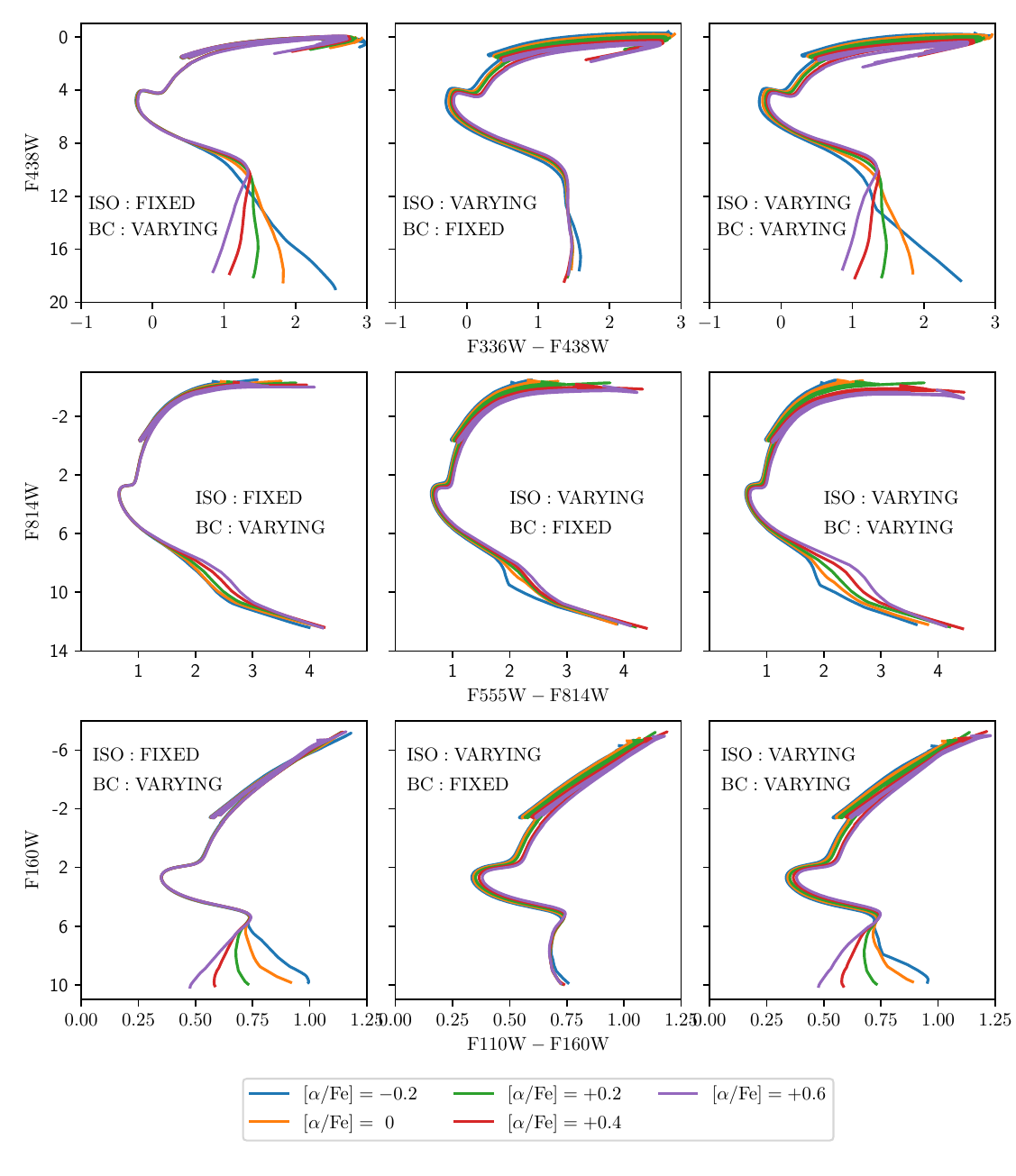}
    \caption{Isochrones at 8 Gyr and $[{\rm Fe/H}]=-0.5$ for three different HST-WFC3 CMDs. The panels demonstrate the effect of varying [$\alpha$/Fe] on the bolometric corrections and the underlying isochrones. The left column shows the effect of varying [$\alpha$/Fe] in the BCs while keeping the underlying isochrone fixed at [$\alpha$/Fe] = +0.2.  The middle panel shows the effect of varying [$\alpha$/Fe] in the isochrones while leaving the BCs fixed at [$\alpha$/Fe] = +0.2.  The right panel shows the combined effect of varying both components together.}
    \label{fig:alphaCMD}
\end{figure*}

Figure \ref{fig:alphaCMD} demonstrates the cumulative effects of $\alpha$-enhancement on stellar models in the Hertzsprung-Russell diagram and the CMD. The HST/WFC3 photometric system is used because it spans the UV, optical, and 
near-infrared.  Figure \ref{fig:alphaCMD} shows the same isochrones with fixed age of 8 Gyr and fixed $[{\rm Fe/H}]=-0.5$. Each row represents a different CMD from the UV (top), to the optical (middle), to the near-infrared (bottom). The models in the left column show the case where the underlying isochrone is kept fixed --- in this case we use the isochrone with [$\alpha$/Fe]=$+0.2$ --- and the level of $\alpha$-enhancement varies from $-0.2$ to $+0.6$ only in the bolometric corrections, thereby isolating the spectral effect of $\alpha$-enhancement. The middle column of Figure \ref{fig:alphaCMD} shows the converse: in this case [$\alpha$/Fe] varies in the underlying isochrones (as in Figure \ref{fig:alpha_HRD_all}) but is kept fixed at [$\alpha$/Fe]=$+0.2$ in the bolometric corrections. The right column of Figure \ref{fig:alphaCMD} shows the combined effects of $\alpha$-enhancement on both the underlying isochrones and in the bolometric corrections.

The most dramatic effect in Figure \ref{fig:alphaCMD} is seen below the ``knee'' in the CMD, where the effect of molecular absorption in the atmosphere becomes prominent.  In this regime, the formation of molecules involving C, N, and O is modified by variations to the oxygen abundance via [$\alpha$/Fe].  In the near-infrared CMD, it is the formation of water molecules and the increased absorption primarily in the F160W filter.  In the ultraviolet CMD, the oxygen competes with carbon and nitrogen for hydrogen atoms to a varying extent, depending on [$\alpha$/Fe] \citep[see ][]{Sbordone2011,Dotter2015}.  The variations seen in the optical CMD are less pronounced due to the lack of molecular lines in these filters.


\subsection{Comparisons with other models}

\begin{figure}
    \centering
    \includegraphics{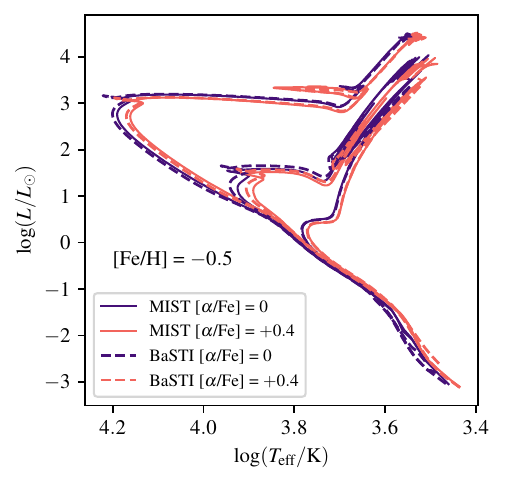}
    \caption{Isochrones at 100~Myr, 1~Gyr, and 10~Gyr from MIST (solid lines) and BaSTI (dashed lines). Darker colors show solar-scaled $\afe = 0$ isochrones, while lighter colors show $\afe = +0.4$ enhanced isochrones.}
    \label{fig:BaSTI_comparison}
\end{figure}

Figure~\ref{fig:BaSTI_comparison} shows MIST $\alpha$-enhanced isochrones (solid lines) compared to $\alpha$-enhanced isochrones from BaSTI \citep[dashed lines][]{2021ApJ...908..102P} at the same [Fe/H]. Considering that Figure \ref{fig:BaSTI_comparison} is based on models coming from two different stellar evolution codes with differences in the adopted physics, the overall agreement is reassuring.


\subsection{Comparisons with observations}\label{sec:data}

\begin{figure*}
    \centering
    \includegraphics[width=\textwidth]{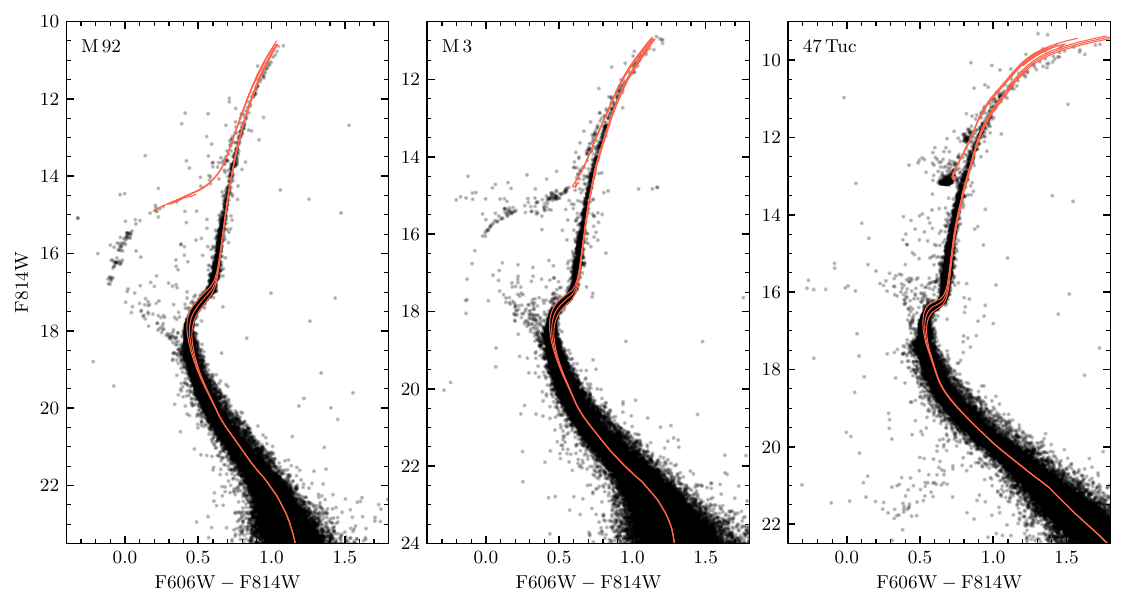}
    \caption{Comparison of metal-poor, $\alpha$-enhanced isochrones with the HST/ACS CMDs of three Galactic globular clusters: M\,92 (left), M\,3 (center), and 47\,Tuc (right). Four ages from 10 to 13 Gyr are shown; all have [$\alpha$/Fe]=+0.4.  The data are from the ACS Survey of Galactic Globular Clusters \citep{Sarajedini07,Anderson08}. Model [Fe/H], distance modulus, and reddening parameters are given in Table \ref{tab:GCs}.        }
    \label{fig:GCs}
\end{figure*}

\begin{figure*}
    \centering
    \includegraphics[width=\textwidth]{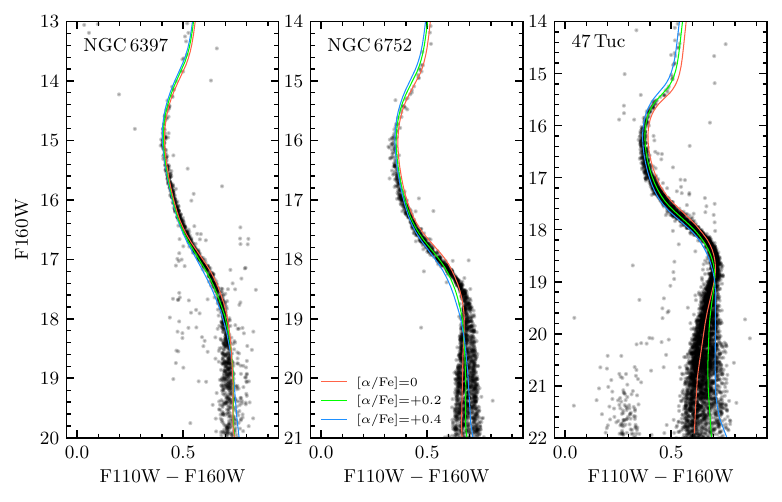}
    \caption{Comparison of metal-poor isochrones with the HST/WFC3 CMDs of three Galactic globular clusters: NGC\,6397 (left), NGC\,6752 (center), and 47\,Tuc (right). In each panel, the observations are compared with models at a fixed [Fe/H] and age but with three different values of [$\alpha$/Fe].  Model [Fe/H], distance modulus, and reddening parameters are given in Table \ref{tab:GCs}. }
    \label{fig:GCsIR}
\end{figure*}

\begin{table}
\centering
\begin{longtable}{l l l l} \\ 
\caption{Globular cluster and model parameters for Figures \ref{fig:GCs} and \ref{fig:GCsIR}} \\
\hline
\hline \noalign{\smallskip}
Cluster & DM & E(B$-$V) & [Fe/H] \\
\hline 
47\,Tuc & 13.266 & 0.04 & $-0.72$ \\
M\,3 & 15.043 & 0.01 & -1.50 \\
M\,92 & 14.595 & 0.02 & -2.31 \\
NGC\,6497 & 11.920 & 0.18 & -2.02 \\ 
NGC\,6752 & 13.010 & 0.04 & -1.54 \\
\hline \hline
   \label{tab:GCs}
\end{longtable}
\tablecomments{Distance modulus (DM) and reddening (E(B$-$V)) from \citet{refId0}; [Fe/H] from the latest edition of the \citet{harris96} catalog.}
\end{table}

In Figures \ref{fig:GCs} and \ref{fig:GCsIR} we compare MIST $\alpha$-enhanced isochrones with CMDs of three Galactic globular clusters.  These comparisons are not meant to be a detailed analysis of these clusters; we merely wish to demonstrate the capabilities of MIST. The isochrones were interpolated in [Fe/H] from our grid.  Evolution past the AGB is not included in either Figure \ref{fig:GCs} or \ref{fig:GCsIR} for clarity.  We note that while globular clusters are representative of $\alpha$-enhanced stellar populations, their stars also---and perhaps more importantly---contains variations in the light elements that are not consistent with $\alpha$-enhancement \citep[cf.][]{2022Univ....8..359M}.

The data shown in Figure \ref{fig:GCs} were taken from the ACS Survey of Galactic Globular Clusters \citep{Sarajedini07,Anderson08} and we have chosen clusters spanning approximately 1.5 dex in [Fe/H].  Details of model parameters, distance moduli, and extinction parameters involved are provided in Table \ref{tab:GCs}. The figures shows 4 isochrones in each panel, these are 10, 11, 12, and 13 Gyr, respectively, and all have [$\alpha$/Fe]=+0.4.  Note that MIST isochrones include only a single value for RGB mass-loss and so will not describe a full zero-age horizontal branch sequence. Hence one should not expect to see full coverage of the horizontal branch stars in any CMD.  

Figure \ref{fig:GCsIR} shows WFC3/IR photometry of NGC\,6397 \citep{2022Univ....8..359M}, NGC\,6752 \citep{2022ApJ...927..207D}, and 47\,Tuc \citep{2025A&A...698A.247M}.  In contrast to Figure \ref{fig:GCs}, we show only one age per panel (13 Gyr) but vary [$\alpha$/Fe] to show the marked effect of varying the $\alpha$-capture elements on the lower main sequence, as mentioned in $\S$\ref{sec:sensitivity}.  Again the parameters used for the figure are given in Table \ref{tab:GCs}.


\section{Summary}
The MIST isochrones have been updated and expanded to include the effects of $\alpha$-enhancement.  The $\alpha$-enhanced grid covers the range of $-3 \leq$ [Fe/H] $\leq +0.5$ and includes 
[$\alpha$/Fe]=$-0.2$, 0, +0.2, +0.4, and +0.6 for all [Fe/H] values, excepting [Fe/H]=+0.5, [$\alpha$/Fe]=+0.6 due to lack of input physics above $Z$=0.1.  Stellar evolution tracks and isochrones are provided from the MIST project webpage.  The file formats are identical to those used in Paper 1.

A number of relatively minor changes to the physics used in \texttt{MESA} have been made but the resulting models have been validated with comparisons to models from Paper 1 and BaSTI as well as observations of Galactic globular clusters.  The necessary tools to reproduce the models presented in this paper are made available via Zenodo;\footnote{\url{https://zenodo.org/records/15232687}, \url{https://zenodo.org/records/15213406}} links to these repositories are included on the MIST project webpage.

\begin{acknowledgments}
The work presented in this paper would not have been possible without the tremendous contributions of Bill Paxton.  Thanks to Bill and all of the \texttt{MESA} developers for creating and supporting \texttt{MESA}.  
AD received financial support from HST-AR-15793. Support for Program number HST-AR-15793 was provided by NASA through a grant from the Space Telescope Science Institute, which is operated by the Association of Universities for Research in Astronomy, Incorporated, under NASA contract NAS5-26555.  AD also received financial support from the European Research Council (ERC) under the Horizon Europe programme (Synergy Grant agreement No. 101071505: 4D-STAR). This work is partially funded by the European Union. Views and opinions expressed are, however, those of the authors only and do not necessarily reflect those of the European Union or the European Research Council. The Flatiron Institute is supported by the Simons Foundation.     
\end{acknowledgments}

\begin{appendix}
Section \ref{sec:MESA} mentions that a few changes have been introduced into \texttt{MESA} revision 11701 for this paper. The purpose of this Appendix is to catalog the changes made to source code files.  All files that are mentioned here, as well as all data and input files required to reproduce the models, are included in the Zenodo repositories \citep{MIST_11701, MESA_work}. The changes are listed in no particular order.
\begin{itemize}
    \item In SVN commit 13223 the mass of the proton (\texttt{mp}) was changed to the atomic mass unit (\texttt{amu}) at line number 283 in \texttt{mesa/eos/private/eospc\_eval.f90}. The result of this change is to smooth out the transition between the PC and HELM equations of state.
    \item FreeEOS tables were imported from \citet{2023ApJS..265...15J}. This required changing parameters in the file \texttt{mesa/eos/public/eos\_def.f} related to the number of $X$ and $Z$ values supported by the EOS tables.
    \item The file \texttt{mesa/atm/private/table\_atm.f90} was modified to take into account the [$\alpha$/Fe] value listed in the table headers.  The [$\alpha$/Fe] value was included in the table header specification since \citet{Paxton11} but was never utilized because all previous versions of the atmosphere boundary condition tables were only for scaled-solar composition.  The atmosphere boundary condition tables used in this paper were calculated with $\alpha$-enhanced chemical compositions.
    \item The file \texttt{mesa/star/private/turbulent\_diffusion.f90} was modified to include the turbulent diffusion model described by \citet{Dotter2017}. 
    \item In SVN commit 11971 the file \texttt{mesa/star/private/element\_diffusion.f90} was changed at lines 184-185 to fix a bug in identifying the location of a convection zone.
\end{itemize}
\end{appendix}

\software{ MESA SDK \citep{MESASDK}, MESA rev.\ 11701 \citep{MESA_11701}, iso \citep{2016ascl.soft01021D}, matplotlib \citep{matplotlib}, numpy \citep{numpy}, Python \citep{python3}}

\clearpage
\bibliography{bibliography}{}

@ARTICLE{Joyce2024,
       author = {{Joyce}, Meridith and {Moln{\'a}r}, L{\'a}szl{\'o} and {Cinquegrana}, Giulia and {Karakas}, Amanda and {Tayar}, Jamie and {Tarczay-Neh{\'e}z}, D{\'o}ra},
        title = "{Stellar Evolution in Real Time. II. R Hydrae and an Open-Source Grid of >3000 Seismic TP-AGB Models Computed with MESA}",
      journal = {\apj},
         year = 2024,
        month = aug,
       volume = {971},
       number = {2},
          eid = {186},
        pages = {186},
          doi = {10.3847/1538-4357/ad534a},
archivePrefix = {arXiv},
       eprint = {2401.16142},
 primaryClass = {astro-ph.SR},
       adsurl = {https://ui.adsabs.harvard.edu/abs/2024ApJ...971..186J}
}

@ARTICLE{Bensby2014,
       author = {{Bensby}, T. and {Feltzing}, S. and {Oey}, M.~S.},
        title = "{Exploring the Milky Way stellar disk. A detailed elemental abundance study of 714 F and G dwarf stars in the solar neighbourhood}",
      journal = {\aap},
         year = 2014,
        month = feb,
       volume = {562},
          eid = {A71},
        pages = {A71},
          doi = {10.1051/0004-6361/201322631},
archivePrefix = {arXiv},
       eprint = {1309.2631},
 primaryClass = {astro-ph.GA},
       adsurl = {https://ui.adsabs.harvard.edu/abs/2014A&A...562A..71B}
}

@software{2016ascl.soft01021D,
       author = {{Dotter}, Aaron},
        title = "{ISO: Isochrone construction}",
 howpublished = {Astrophysics Source Code Library, record ascl:1601.021},
         year = 2016,
        month = jan,
          eid = {ascl:1601.021},
       adsurl = {https://ui.adsabs.harvard.edu/abs/2016ascl.soft01021D}
}

@ARTICLE{1986HiA.....7..833K,
       author = {{Koornneef}, J. and {Bohlin}, R. and {Buser}, R. and {Horne}, K. and {Turnshek}, D.},
        title = "{Synthetic photometry and the calibration of the Hubble Space Telescope.}",
      journal = {Highlights of Astronomy},
         year = 1986,
        month = jan,
       volume = {7},
        pages = {833-843},
       adsurl = {https://ui.adsabs.harvard.edu/abs/1986HiA.....7..833K}
}

@ARTICLE{1974ApJS...27...21O,
       author = {{Oke}, J.~B.},
        title = "{Absolute Spectral Energy Distributions for White Dwarfs}",
      journal = {\apjs},
         year = 1974,
        month = feb,
       volume = {27},
        pages = {21},
          doi = {10.1086/190287},
       adsurl = {https://ui.adsabs.harvard.edu/abs/1974ApJS...27...21O}
}

@ARTICLE{Sbordone2011,
       author = {{Sbordone}, L. and {Salaris}, M. and {Weiss}, A. and {Cassisi}, S.},
        title = "{Photometric signatures of multiple stellar populations in Galactic globular clusters}",
      journal = {\aap},
         year = 2011,
        month = oct,
       volume = {534},
          eid = {A9},
        pages = {A9},
          doi = {10.1051/0004-6361/201116714},
archivePrefix = {arXiv},
       eprint = {1103.5863},
 primaryClass = {astro-ph.SR},
       adsurl = {https://ui.adsabs.harvard.edu/abs/2011A&A...534A...9S}
}

@ARTICLE{Marigo2017,
       author = {{Marigo}, Paola and {Girardi}, L{\'e}o and {Bressan}, Alessandro and {Rosenfield}, Philip and {Aringer}, Bernhard and {Chen}, Yang and {Dussin}, Marco and {Nanni}, Ambra and {Pastorelli}, Giada and {Rodrigues}, Tha{\'\i}se S. and {Trabucchi}, Michele and {Bladh}, Sara and {Dalcanton}, Julianne and {Groenewegen}, Martin A.~T. and {Montalb{\'a}n}, Josefina and {Wood}, Peter R.},
        title = "{A New Generation of PARSEC-COLIBRI Stellar Isochrones Including the TP-AGB Phase}",
      journal = {\apj},
         year = 2017,
        month = jan,
       volume = {835},
       number = {1},
          eid = {77},
        pages = {77},
          doi = {10.3847/1538-4357/835/1/77},
archivePrefix = {arXiv},
       eprint = {1701.08510},
 primaryClass = {astro-ph.SR},
       adsurl = {https://ui.adsabs.harvard.edu/abs/2017ApJ...835...77M}
}

@ARTICLE{Salasnich2000,
       author = {{Salasnich}, B. and {Girardi}, L. and {Weiss}, A. and {Chiosi}, C.},
        title = "{Evolutionary tracks and isochrones for alpha -enhanced stars}",
      journal = {\aap},
         year = 2000,
        month = sep,
       volume = {361},
        pages = {1023-1035},
          doi = {10.48550/arXiv.astro-ph/0007388},
archivePrefix = {arXiv},
       eprint = {astro-ph/0007388},
 primaryClass = {astro-ph},
       adsurl = {https://ui.adsabs.harvard.edu/abs/2000A&A...361.1023S}
}

@ARTICLE{Nguyen2022,
       author = {{Nguyen}, C.~T. and {Costa}, G. and {Girardi}, L. and {Volpato}, G. and {Bressan}, A. and {Chen}, Y. and {Marigo}, P. and {Fu}, X. and {Goudfrooij}, P.},
        title = "{PARSEC V2.0: Stellar tracks and isochrones of low- and intermediate-mass stars with rotation}",
      journal = {\aap},
         year = 2022,
        month = sep,
       volume = {665},
          eid = {A126},
        pages = {A126},
          doi = {10.1051/0004-6361/202244166},
archivePrefix = {arXiv},
       eprint = {2207.08642},
 primaryClass = {astro-ph.SR},
       adsurl = {https://ui.adsabs.harvard.edu/abs/2022A&A...665A.126N}
}

@ARTICLE{2025A&A...698A.247M,
       author = {{Milone}, A.~P. and {Marino}, A.~F. and {Bernizzoni}, M. and {Muratore}, F. and {Legnardi}, M.~V. and {Barbieri}, M. and {Bortolan}, E. and {Bouras}, A. and {Bruce}, J. and {Cordoni}, G. and {D'Antona}, F. and {Dell'Agli}, F. and {Dondoglio}, E. and {Grimaldi}, I.~M. and {Jang}, S. and {Lagioia}, E.~P. and {Lee}, J.-W. and {Lionetto}, S. and {Mohandasan}, A. and {Pang}, X. and {Pianta}, C. and {Posenato}, M. and {Renzini}, A. and {Tailo}, M. and {Ventura}, C. and {Ventura}, P. and {Vesperini}, E. and {Ziliotto}, T.},
        title = "{A JWST project on 47 Tucanae: Binaries among multiple populations}",
      journal = {\aap},
         year = 2025,
        month = jun,
       volume = {698},
          eid = {A247},
        pages = {A247},
          doi = {10.1051/0004-6361/202452136},
archivePrefix = {arXiv},
       eprint = {2503.19214},
 primaryClass = {astro-ph.SR},
       adsurl = {https://ui.adsabs.harvard.edu/abs/2025A&A...698A.247M}
}

@ARTICLE{2022ApJ...927..207D,
       author = {{Dondoglio}, E. and {Milone}, A.~P. and {Renzini}, A. and {Vesperini}, E. and {Lagioia}, E.~P. and {Marino}, A.~F. and {Bellini}, A. and {Carlos}, M. and {Cordoni}, G. and {Jang}, S. and {Legnardi}, M.~V. and {Libralato}, M. and {Mohandasan}, A. and {D'Antona}, F. and {Martorano}, M. and {Muratore}, F. and {Tailo}, M.},
        title = "{Survey of Multiple Populations in Globular Clusters among Very-low-mass Stars}",
      journal = {\apj},
         year = 2022,
        month = mar,
       volume = {927},
       number = {2},
          eid = {207},
        pages = {207},
          doi = {10.3847/1538-4357/ac5046},
archivePrefix = {arXiv},
       eprint = {2201.08631},
 primaryClass = {astro-ph.SR},
       adsurl = {https://ui.adsabs.harvard.edu/abs/2022ApJ...927..207D}
}

@ARTICLE{2022Univ....8..359M,
       author = {{Milone}, Antonino P. and {Marino}, Anna F.},
        title = "{Multiple Populations in Star Clusters}",
      journal = {Universe},
         year = 2022,
        month = jun,
       volume = {8},
       number = {7},
          eid = {359},
        pages = {359},
          doi = {10.3390/universe8070359},
archivePrefix = {arXiv},
       eprint = {2206.10564},
 primaryClass = {astro-ph.GA},
       adsurl = {https://ui.adsabs.harvard.edu/abs/2022Univ....8..359M}
}

@article{ refId0,
	author = {{Gaia Collaboration} and {Babusiaux, C.} and {van Leeuwen, F.} and {Barstow, M. A.} and {Jordi, C.} and {Vallenari, A.} and {Bossini, D.} and {Bressan, A.} and {Cantat-Gaudin, T.} and {van Leeuwen, M.} and {Brown, A. G. A.} and {Prusti, T.} and {de Bruijne, J. H. J.} and {Bailer-Jones, C. A. L.} and {Biermann, M.} and {Evans, D. W.} and {Eyer, L.} and {Jansen, F.} and {Klioner, S. A.} and {Lammers, U.} and {Lindegren, L.} and {Luri, X.} and {Mignard, F.} and {Panem, C.} and {Pourbaix, D.} and {Randich, S.} and {Sartoretti, P.} and {Siddiqui, H. I.} and {Soubiran, C.} and {Walton, N. A.} and {Arenou, F.} and {Bastian, U.} and {Cropper, M.} and {Drimmel, R.} and {Katz, D.} and {Lattanzi, M. G.} and {Bakker, J.} and {Cacciari, C.} and {Castañeda, J.} and {Chaoul, L.} and {Cheek, N.} and {De Angeli, F.} and {Fabricius, C.} and {Guerra, R.} and {Holl, B.} and {Masana, E.} and {Messineo, R.} and {Mowlavi, N.} and {Nienartowicz, K.} and {Panuzzo, P.} and {Portell, J.} and {Riello, M.} and {Seabroke, G. M.} and {Tanga, P.} and {Thévenin, F.} and {Gracia-Abril, G.} and {Comoretto, G.} and {Garcia-Reinaldos, M.} and {Teyssier, D.} and {Altmann, M.} and {Andrae, R.} and {Audard, M.} and {Bellas-Velidis, I.} and {Benson, K.} and {Berthier, J.} and {Blomme, R.} and {Burgess, P.} and {Busso, G.} and {Carry, B.} and {Cellino, A.} and {Clementini, G.} and {Clotet, M.} and {Creevey, O.} and {Davidson, M.} and {De Ridder, J.} and {Delchambre, L.} and {Dell’Oro, A.} and {Ducourant, C.} and {Fernández-Hernández, J.} and {Fouesneau, M.} and {Frémat, Y.} and {Galluccio, L.} and {García-Torres, M.} and {González-Núñez, J.} and {González-Vidal, J. J.} and {Gosset, E.} and {Guy, L. P.} and {Halbwachs, J.-L.} and {Hambly, N. C.} and {Harrison, D. L.} and {Hernández, J.} and {Hestroffer, D.} and {Hodgkin, S. T.} and {Hutton, A.} and {Jasniewicz, G.} and {Jean-Antoine-Piccolo, A.} and {Jordan, S.} and {Korn, A. J.} and {Krone-Martins, A.} and {Lanzafame, A. C.} and {Lebzelter, T.} and {Löffler, W.} and {Manteiga, M.} and {Marrese, P. M.} and {Martín-Fleitas, J. M.} and {Moitinho, A.} and {Mora, A.} and {Muinonen, K.} and {Osinde, J.} and {Pancino, E.} and {Pauwels, T.} and {Petit, J.-M.} and {Recio-Blanco, A.} and {Richards, P. J.} and {Rimoldini, L.} and {Robin, A. C.} and {Sarro, L. M.} and {Siopis, C.} and {Smith, M.} and {Sozzetti, A.} and {Süveges, M.} and {Torra, J.} and {van Reeven, W.} and {Abbas, U.} and {Abreu Aramburu, A.} and {Accart, S.} and {Aerts, C.} and {Altavilla, G.} and {Álvarez, M. A.} and {Alvarez, R.} and {Alves, J.} and {Anderson, R. I.} and {Andrei, A. H.} and {Anglada Varela, E.} and {Antiche, E.} and {Antoja, T.} and {Arcay, B.} and {Astraatmadja, T. L.} and {Bach, N.} and {Baker, S. G.} and {Balaguer-Núñez, L.} and {Balm, P.} and {Barache, C.} and {Barata, C.} and {Barbato, D.} and {Barblan, F.} and {Barklem, P. S.} and {Barrado, D.} and {Barros, M.} and {Bartholomé Muñoz, L.} and {Bassilana, J.-L.} and {Becciani, U.} and {Bellazzini, M.} and {Berihuete, A.} and {Bertone, S.} and {Bianchi, L.} and {Bienaymé, O.} and {Blanco-Cuaresma, S.} and {Boch, T.} and {Boeche, C.} and {Bombrun, A.} and {Borrachero, R.} and {Bouquillon, S.} and {Bourda, G.} and {Bragaglia, A.} and {Bramante, L.} and {Breddels, M. A.} and {Brouillet, N.} and {Brüsemeister, T.} and {Brugaletta, E.} and {Bucciarelli, B.} and {Burlacu, A.} and {Busonero, D.} and {Butkevich, A. G.} and {Buzzi, R.} and {Caffau, E.} and {Cancelliere, R.} and {Cannizzaro, G.} and {Carballo, R.} and {Carlucci, T.} and {Carrasco, J. M.} and {Casamiquela, L.} and {Castellani, M.} and {Castro-Ginard, A.} and {Charlot, P.} and {Chemin, L.} and {Chiavassa, A.} and {Cocozza, G.} and {Costigan, G.} and {Cowell, S.} and {Crifo, F.} and {Crosta, M.} and {Crowley, C.} and {Cuypers, J.} and {Dafonte, C.} and {Damerdji, Y.} and {Dapergolas, A.} and {David, P.} and {David, M.} and {de Laverny, P.} and {De Luise, F.} and {De March, R.} and {de Martino, D.} and {de Souza, R.} and {de Torres, A.} and {Debosscher, J.} and {del Pozo, E.} and {Delbo, M.} and {Delgado, A.} and {Delgado, H. E.} and {Diakite, S.} and {Diener, C.} and {Distefano, E.} and {Dolding, C.} and {Drazinos, P.} and {Durán, J.} and {Edvardsson, B.} and {Enke, H.} and {Eriksson, K.} and {Esquej, P.} and {Eynard Bontemps, G.} and {Fabre, C.} and {Fabrizio, M.} and {Faigler, S.} and {Falcão, A. J.} and {Farràs Casas, M.} and {Federici, L.} and {Fedorets, G.} and {Fernique, P.} and {Figueras, F.} and {Filippi, F.} and {Findeisen, K.} and {Fonti, A.} and {Fraile, E.} and {Fraser, M.} and {Frézouls, B.} and {Gai, M.} and {Galleti, S.} and {Garabato, D.} and {García-Sedano, F.} and {Garofalo, A.} and {Garralda, N.} and {Gavel, A.} and {Gavras, P.} and {Gerssen, J.} and {Geyer, R.} and {Giacobbe, P.} and {Gilmore, G.} and {Girona, S.} and {Giuffrida, G.} and {Glass, F.} and {Gomes, M.} and {Granvik, M.} and {Gueguen, A.} and {Guerrier, A.} and {Guiraud, J.} and {Gutié, R.} and {Haigron, R.} and {Hatzidimitriou, D.} and {Hauser, M.} and {Haywood, M.} and {Heiter, U.} and {Helmi, A.} and {Heu, J.} and {Hilger, T.} and {Hobbs, D.} and {Hofmann, W.} and {Holland, G.} and {Huckle, H. E.} and {Hypki, A.} and {Icardi, V.} and {Janßen, K.} and {Jevardat de Fombelle, G.} and {Jonker, P. G.} and {Juhász, Á. L.} and {Julbe, F.} and {Karampelas, A.} and {Kewley, A.} and {Klar, J.} and {Kochoska, A.} and {Kohley, R.} and {Kolenberg, K.} and {Kontizas, M.} and {Kontizas, E.} and {Koposov, S. E.} and {Kordopatis, G.} and {Kostrzewa-Rutkowska, Z.} and {Koubsky, P.} and {Lambert, S.} and {Lanza, A. F.} and {Lasne, Y.} and {Lavigne, J.-B.} and {Le Fustec, Y.} and {Le Poncin-Lafitte, C.} and {Lebreton, Y.} and {Leccia, S.} and {Leclerc, N.} and {Lecoeur-Taibi, I.} and {Lenhardt, H.} and {Leroux, F.} and {Liao, S.} and {Licata, E.} and {Lindstrøm, H. E. P.} and {Lister, T. A.} and {Livanou, E.} and {Lobel, A.} and {López, M.} and {Managau, S.} and {Mann, R. G.} and {Mantelet, G.} and {Marchal, O.} and {Marchant, J. M.} and {Marconi, M.} and {Marinoni, S.} and {Marschalkó, G.} and {Marshall, D. J.} and {Martino, M.} and {Marton, G.} and {Mary, N.} and {Massari, D.} and {Matijevič, G.} and {Mazeh, T.} and {McMillan, P. J.} and {Messina, S.} and {Michalik, D.} and {Millar, N. R.} and {Molina, D.} and {Molinaro, R.} and {Molnár, L.} and {Montegriffo, P.} and {Mor, R.} and {Morbidelli, R.} and {Morel, T.} and {Morris, D.} and {Mulone, A. F.} and {Muraveva, T.} and {Musella, I.} and {Nelemans, G.} and {Nicastro, L.} and {Noval, L.} and {O’Mullane, W.} and {Ordénovic, C.} and {Ordóñez-Blanco, D.} and {Osborne, P.} and {Pagani, C.} and {Pagano, I.} and {Pailler, F.} and {Palacin, H.} and {Palaversa, L.} and {Panahi, A.} and {Pawlak, M.} and {Piersimoni, A. M.} and {Pineau, F.-X.} and {Plachy, E.} and {Plum, G.} and {Poggio, E.} and {Poujoulet, E.} and {Prša, A.} and {Pulone, L.} and {Racero, E.} and {Ragaini, S.} and {Rambaux, N.} and {Ramos-Lerate, M.} and {Regibo, S.} and {Reylé, C.} and {Riclet, F.} and {Ripepi, V.} and {Riva, A.} and {Rivard, A.} and {Rixon, G.} and {Roegiers, T.} and {Roelens, M.} and {Romero-Gómez, M.} and {Rowell, N.} and {Royer, F.} and {Ruiz-Dern, L.} and {Sadowski, G.} and {Sagristà Sellés, T.} and {Sahlmann, J.} and {Salgado, J.} and {Salguero, E.} and {Sanna, N.} and {Santana-Ros, T.} and {Sarasso, M.} and {Savietto, H.} and {Schultheis, M.} and {Sciacca, E.} and {Segol, M.} and {Segovia, J. C.} and {Ségransan, D.} and {Shih, I-C.} and {Siltala, L.} and {Silva, A. F.} and {Smart, R. L.} and {Smith, K. W.} and {Solano, E.} and {Solitro, F.} and {Sordo, R.} and {Soria Nieto, S.} and {Souchay, J.} and {Spagna, A.} and {Spoto, F.} and {Stampa, U.} and {Steele, I. A.} and {Steidelmüller, H.} and {Stephenson, C. A.} and {Stoev, H.} and {Suess, F. F.} and {Surdej, J.} and {Szabados, L.} and {Szegedi-Elek, E.} and {Tapiador, D.} and {Taris, F.} and {Tauran, G.} and {Taylor, M. B.} and {Teixeira, R.} and {Terrett, D.} and {Teyssandier, P.} and {Thuillot, W.} and {Titarenko, A.} and {Torra Clotet, F.} and {Turon, C.} and {Ulla, A.} and {Utrilla, E.} and {Uzzi, S.} and {Vaillant, M.} and {Valentini, G.} and {Valette, V.} and {van Elteren, A.} and {Van Hemelryck, E.} and {Vaschetto, M.} and {Vecchiato, A.} and {Veljanoski, J.} and {Viala, Y.} and {Vicente, D.} and {Vogt, S.} and {von Essen, C.} and {Voss, H.} and {Votruba, V.} and {Voutsinas, S.} and {Walmsley, G.} and {Weiler, M.} and {Wertz, O.} and {Wevers, T.} and {Wyrzykowski, Ł.} and {Yoldas, A.} and {Žerjal, M.} and {Ziaeepour, H.} and {Zorec, J.} and {Zschocke, S.} and {Zucker, S.} and {Zurbach, C.} and {Zwitter, and T.}},
	title = {Gaia Data Release 2 - Observational Hertzsprung-Russell diagrams★},
	DOI= "10.1051/0004-6361/201832843",
	url= "https://doi.org/10.1051/0004-6361/201832843",
	journal = {\aap},
	year = 2018,
	volume = 616,
	pages = "A10",
}

@ARTICLE{Mosser2012,
       author = {{Mosser}, B. and {Goupil}, M.~J. and {Belkacem}, K. and {Marques}, J.~P. and {Beck}, P.~G. and {Bloemen}, S. and {De Ridder}, J. and {Barban}, C. and {Deheuvels}, S. and {Elsworth}, Y. and {Hekker}, S. and {Kallinger}, T. and {Ouazzani}, R.~M. and {Pinsonneault}, M. and {Samadi}, R. and {Stello}, D. and {Garc{\'\i}a}, R.~A. and {Klaus}, T.~C. and {Li}, J. and {Mathur}, S. and {Morris}, R.~L.},
        title = "{Spin down of the core rotation in red giants}",
      journal = {\aap},
         year = 2012,
        month = dec,
       volume = {548},
          eid = {A10},
        pages = {A10},
          doi = {10.1051/0004-6361/201220106},
archivePrefix = {arXiv},
       eprint = {1209.3336},
 primaryClass = {astro-ph.SR},
       adsurl = {https://ui.adsabs.harvard.edu/abs/2012A&A...548A..10M}
}

@ARTICLE{Gehan2018,
       author = {{Gehan}, C. and {Mosser}, B. and {Michel}, E. and {Samadi}, R. and {Kallinger}, T.},
        title = "{Core rotation braking on the red giant branch for various mass ranges}",
      journal = {\aap},
         year = 2018,
        month = aug,
       volume = {616},
          eid = {A24},
        pages = {A24},
          doi = {10.1051/0004-6361/201832822},
archivePrefix = {arXiv},
       eprint = {1802.04558},
 primaryClass = {astro-ph.SR},
       adsurl = {https://ui.adsabs.harvard.edu/abs/2018A&A...616A..24G}
}

@ARTICLE{Deheuvels2015,
       author = {{Deheuvels}, S. and {Ballot}, J. and {Beck}, P.~G. and {Mosser}, B. and {{\O}stensen}, R. and {Garc{\'\i}a}, R.~A. and {Goupil}, M.~J.},
        title = "{Seismic evidence for a weak radial differential rotation in intermediate-mass core helium burning stars}",
      journal = {\aap},
         year = 2015,
        month = aug,
       volume = {580},
          eid = {A96},
        pages = {A96},
          doi = {10.1051/0004-6361/201526449},
archivePrefix = {arXiv},
       eprint = {1506.02704},
 primaryClass = {astro-ph.SR},
       adsurl = {https://ui.adsabs.harvard.edu/abs/2015A&A...580A..96D}
}

@ARTICLE{Hermes2017,
       author = {{Hermes}, J.~J. and {G{\"a}nsicke}, B.~T. and {Kawaler}, Steven D. and {Greiss}, S. and {Tremblay}, P. -E. and {Gentile Fusillo}, N.~P. and {Raddi}, R. and {Fanale}, S.~M. and {Bell}, Keaton J. and {Dennihy}, E. and {Fuchs}, J.~T. and {Dunlap}, B.~H. and {Clemens}, J.~C. and {Montgomery}, M.~H. and {Winget}, D.~E. and {Chote}, P. and {Marsh}, T.~R. and {Redfield}, S.},
        title = "{White Dwarf Rotation as a Function of Mass and a Dichotomy of Mode Line Widths: Kepler Observations of 27 Pulsating DA White Dwarfs through K2 Campaign 8}",
      journal = {\apjs},
         year = 2017,
        month = oct,
       volume = {232},
       number = {2},
          eid = {23},
        pages = {23},
          doi = {10.3847/1538-4365/aa8bb5},
archivePrefix = {arXiv},
       eprint = {1709.07004},
 primaryClass = {astro-ph.SR},
       adsurl = {https://ui.adsabs.harvard.edu/abs/2017ApJS..232...23H}
}

@ARTICLE{Fuller2019,
       author = {{Fuller}, Jim and {Piro}, Anthony L. and {Jermyn}, Adam S.},
        title = "{Slowing the spins of stellar cores}",
      journal = {\mnras},
         year = 2019,
        month = may,
       volume = {485},
       number = {3},
        pages = {3661-3680},
          doi = {10.1093/mnras/stz514},
archivePrefix = {arXiv},
       eprint = {1902.08227},
 primaryClass = {astro-ph.SR},
       adsurl = {https://ui.adsabs.harvard.edu/abs/2019MNRAS.485.3661F}
}

@BOOK{1984psen.book.....C,
       author = {{Clayton}, D.~D.},
        title = "{Principles of stellar evolution and nucleosynthesis.}",
         year = 1984,
       adsurl = {https://ui.adsabs.harvard.edu/abs/1984psen.book.....C}
}

@BOOK{Kopal1978,
       author = {{Kopal}, Z.},
        title = "{Dynamics of close binary systems}",
         year = 1978,
          doi = {10.1007/978-94-009-9780-6},
       adsurl = {https://ui.adsabs.harvard.edu/abs/1978ASSL...68.....K}
}

@ARTICLE{2011A&A...533A..43E,
       author = {{Espinosa Lara}, F. and {Rieutord}, M.},
        title = "{Gravity darkening in rotating stars}",
      journal = {\aap},
         year = 2011,
        month = sep,
       volume = {533},
          eid = {A43},
        pages = {A43},
          doi = {10.1051/0004-6361/201117252},
archivePrefix = {arXiv},
       eprint = {1109.3038},
 primaryClass = {astro-ph.SR},
       adsurl = {https://ui.adsabs.harvard.edu/abs/2011A&A...533A..43E}
}

@ARTICLE{2023ApJ...950..115B,
       author = {{Bauer}, Evan B.},
        title = "{Carbon-Oxygen Phase Separation in Modules for Experiments in Stellar Astrophysics (MESA) White Dwarf Models}",
      journal = {\apj},
         year = 2023,
        month = jun,
       volume = {950},
       number = {2},
          eid = {115},
        pages = {115},
          doi = {10.3847/1538-4357/acd057},
archivePrefix = {arXiv},
       eprint = {2303.10110},
 primaryClass = {astro-ph.SR},
       adsurl = {https://ui.adsabs.harvard.edu/abs/2023ApJ...950..115B}
}

@ARTICLE{2007ApJ...663..320F,
       author = {{Fitzpatrick}, E.~L. and {Massa}, D.},
        title = "{An Analysis of the Shapes of Interstellar Extinction Curves. V. The IR-through-UV Curve Morphology}",
      journal = {\apj},
         year = 2007,
        month = jul,
       volume = {663},
       number = {1},
        pages = {320-341},
          doi = {10.1086/518158},
archivePrefix = {arXiv},
       eprint = {0705.0154},
 primaryClass = {astro-ph},
       adsurl = {https://ui.adsabs.harvard.edu/abs/2007ApJ...663..320F}
}

@ARTICLE{2023ApJS..265...15J,
       author = {{Jermyn}, Adam S. and {Bauer}, Evan B. and {Schwab}, Josiah and {Farmer}, R. and {Ball}, Warrick H. and {Bellinger}, Earl P. and {Dotter}, Aaron and {Joyce}, Meridith and {Marchant}, Pablo and {Mombarg}, Joey S.~G. and {Wolf}, William M. and {Sunny Wong}, Tin Long and {Cinquegrana}, Giulia C. and {Farrell}, Eoin and {Smolec}, R. and {Thoul}, Anne and {Cantiello}, Matteo and {Herwig}, Falk and {Toloza}, Odette and {Bildsten}, Lars and {Townsend}, Richard H.~D. and {Timmes}, F.~X.},
        title = "{Modules for Experiments in Stellar Astrophysics (MESA): Time-dependent Convection, Energy Conservation, Automatic Differentiation, and Infrastructure}",
      journal = {\apjs},
         year = 2023,
        month = mar,
       volume = {265},
       number = {1},
          eid = {15},
        pages = {15},
          doi = {10.3847/1538-4365/acae8d},
archivePrefix = {arXiv},
       eprint = {2208.03651},
 primaryClass = {astro-ph.SR},
       adsurl = {https://ui.adsabs.harvard.edu/abs/2023ApJS..265...15J}
}

@ARTICLE{2022A&A...661A.140M,
       author = {{Magg}, Ekaterina and {Bergemann}, Maria and {Serenelli}, Aldo and {Bautista}, Manuel and {Plez}, Bertrand and {Heiter}, Ulrike and {Gerber}, Jeffrey M. and {Ludwig}, Hans-G{\"u}nter and {Basu}, Sarbani and {Ferguson}, Jason W. and {Gallego}, Helena Carvajal and {Gamrath}, S{\'e}bastien and {Palmeri}, Patrick and {Quinet}, Pascal},
        title = "{Observational constraints on the origin of the elements. IV. Standard composition of the Sun}",
      journal = {\aap},
         year = 2022,
        month = may,
       volume = {661},
          eid = {A140},
        pages = {A140},
          doi = {10.1051/0004-6361/202142971},
archivePrefix = {arXiv},
       eprint = {2203.02255},
 primaryClass = {astro-ph.SR},
       adsurl = {https://ui.adsabs.harvard.edu/abs/2022A&A...661A.140M}
}

@ARTICLE{2023A&A...672L...6P,
       author = {{Pietrow}, A.~G.~M. and {Hoppe}, R. and {Bergemann}, M. and {Calvo}, F.},
        title = "{Solar oxygen abundance using SST/CRISP center-to-limb observations of the O I 7772 {\r{A}} line}",
      journal = {\aap},
         year = 2023,
        month = apr,
       volume = {672},
          eid = {L6},
        pages = {L6},
          doi = {10.1051/0004-6361/202346387},
archivePrefix = {arXiv},
       eprint = {2304.01048},
 primaryClass = {astro-ph.SR},
       adsurl = {https://ui.adsabs.harvard.edu/abs/2023A&A...672L...6P}
}

@ARTICLE{2022RNAAS...6..112D,
       author = {{Dotter}, Aaron},
        title = "{T({\ensuremath{\tau}}) Integration Incorporating Convection}",
      journal = {Research Notes of the American Astronomical Society},
         year = 2022,
        month = jun,
       volume = {6},
       number = {6},
          eid = {112},
        pages = {112},
          doi = {10.3847/2515-5172/ac74c6},
       adsurl = {https://ui.adsabs.harvard.edu/abs/2022RNAAS...6..112D}
}

@software{MESASDK,
  author       = {Richard Townsend},
  title        = {MESA SDK for Linux},
  month        = may,
  year         = 2019,
  publisher    = {Zenodo},
  version      = 20190503,
  doi          = {10.5281/zenodo.2669541},
  url          = {https://doi.org/10.5281/zenodo.2669541}
}

@ARTICLE{2018ApJ...856..125H,
       author = {{Hidalgo}, Sebastian L. and {Pietrinferni}, Adriano and {Cassisi}, Santi and {Salaris}, Maurizio and {Mucciarelli}, Alessio and {Savino}, Alessandro and {Aparicio}, Antonio and {Silva Aguirre}, Victor and {Verma}, Kuldeep},
        title = "{The Updated BaSTI Stellar Evolution Models and Isochrones. I. Solar-scaled Calculations}",
      journal = {\apj},
         year = 2018,
        month = apr,
       volume = {856},
       number = {2},
          eid = {125},
        pages = {125},
          doi = {10.3847/1538-4357/aab158},
archivePrefix = {arXiv},
       eprint = {1802.07319},
 primaryClass = {astro-ph.GA},
       adsurl = {https://ui.adsabs.harvard.edu/abs/2018ApJ...856..125H}
}

@ARTICLE{2021ApJ...908..102P,
       author = {{Pietrinferni}, Adriano and {Hidalgo}, Sebastian and {Cassisi}, Santi and {Salaris}, Maurizio and {Savino}, Alessandro and {Mucciarelli}, Alessio and {Verma}, Kuldeep and {Silva Aguirre}, Victor and {Aparicio}, Antonio and {Ferguson}, Jason W.},
        title = "{Updated BaSTI Stellar Evolution Models and Isochrones. II. {\ensuremath{\alpha}}-enhanced Calculations}",
      journal = {\apj},
         year = 2021,
        month = feb,
       volume = {908},
       number = {1},
          eid = {102},
        pages = {102},
          doi = {10.3847/1538-4357/abd4d5},
archivePrefix = {arXiv},
       eprint = {2012.10085},
 primaryClass = {astro-ph.SR},
       adsurl = {https://ui.adsabs.harvard.edu/abs/2021ApJ...908..102P}
}

@ARTICLE{2018MNRAS.476..496F,
       author = {{Fu}, Xiaoting and {Bressan}, Alessandro and {Marigo}, Paola and {Girardi}, L{\'e}o and {Montalb{\'a}n}, Josefina and {Chen}, Yang and {Nanni}, Ambra},
        title = "{New PARSEC data base of {\ensuremath{\alpha}}-enhanced stellar evolutionary tracks and isochrones - I. Calibration with 47 Tuc (NGC 104) and the improvement on RGB bump}",
      journal = {\mnras},
         year = 2018,
        month = may,
       volume = {476},
       number = {1},
        pages = {496-511},
          doi = {10.1093/mnras/sty235},
archivePrefix = {arXiv},
       eprint = {1801.07137},
 primaryClass = {astro-ph.SR},
       adsurl = {https://ui.adsabs.harvard.edu/abs/2018MNRAS.476..496F}
}

@ARTICLE{1952ApJ...116..317O,
       author = {{Oke}, J.~B. and {Schwarzschild}, M.},
        title = "{Inhomogeneous Stellar Models. I. Models with a Convective Core and a Discontinuity in the Chemical Composition.}",
      journal = {\apj},
         year = 1952,
        month = sep,
       volume = {116},
        pages = {317},
          doi = {10.1086/145615},
       adsurl = {https://ui.adsabs.harvard.edu/abs/1952ApJ...116..317O}
}

@article{REACLIB,
	doi = {10.1088/0067-0049/189/1/240},
	url = {https://doi.org/10.1088/0067-0049/189/1/240},
	year = 2010,
	month = {jun},
	publisher = {American Astronomical Society},
	volume = {189},
	number = {1},
	pages = {240--252},
	author = {Richard H. Cyburt and A. Matthew Amthor and Ryan Ferguson and Zach Meisel and Karl Smith and Scott Warren and Alexander Heger and R. D. Hoffman and Thomas Rauscher and Alexander Sakharuk and Hendrik Schatz and F. K. Thielemann and Michael Wiescher},
	title = {{THE} {JINA} {REACLIB} {DATABASE}: {ITS} {RECENT} {UPDATES} {AND} {IMPACT} {ON} {TYPE}-I X-{RAY} {BURSTS}},
	journal = {The Astrophysical Journal Supplement Series}
}

@ARTICLE{IAU.2015.B3,
       author = {{Pr{\v{s}}a}, Andrej and {Harmanec}, Petr and {Torres}, Guillermo and {Mamajek}, Eric and {Asplund}, Martin and {Capitaine}, Nicole and {Christensen-Dalsgaard}, J{\o}rgen and {Depagne}, {\'E}ric and {Haberreiter}, Margit and {Hekker}, Saskia and {Hilton}, James and {Kopp}, Greg and {Kostov}, Veselin and {Kurtz}, Donald W. and {Laskar}, Jacques and {Mason}, Brian D. and {Milone}, Eugene F. and {Montgomery}, Michele and {Richards}, Mercedes and {Schmutz}, Werner and {Schou}, Jesper and {Stewart}, Susan G.},
        title = "{Nominal Values for Selected Solar and Planetary Quantities: IAU 2015 Resolution B3}",
      journal = {\aj},
         year = 2016,
        month = aug,
       volume = {152},
       number = {2},
          eid = {41},
        pages = {41},
          doi = {10.3847/0004-6256/152/2/41},
archivePrefix = {arXiv},
       eprint = {1605.09788},
 primaryClass = {astro-ph.SR},
       adsurl = {https://ui.adsabs.harvard.edu/abs/2016AJ....152...41P}
}

@ARTICLE{IAU.2015.B2,
       author = {{Mamajek}, E.~E. and {Torres}, G. and {Prsa}, A. and {Harmanec}, P. and {Asplund}, M. and {Bennett}, P.~D. and {Capitaine}, N. and {Christensen-Dalsgaard}, J. and {Depagne}, E. and {Folkner}, W.~M. and {Haberreiter}, M. and {Hekker}, S. and {Hilton}, J.~L. and {Kostov}, V. and {Kurtz}, D.~W. and {Laskar}, J. and {Mason}, B.~D. and {Milone}, E.~F. and {Montgomery}, M.~M. and {Richards}, M.~T. and {Schou}, J. and {Stewart}, S.~G.},
        title = "{IAU 2015 Resolution B2 on Recommended Zero Points for the Absolute and Apparent Bolometric Magnitude Scales}",
      journal = {arXiv e-prints},
         year = 2015,
        month = oct,
          eid = {arXiv:1510.06262},
        pages = {arXiv:1510.06262},
archivePrefix = {arXiv},
       eprint = {1510.06262},
 primaryClass = {astro-ph.SR},
       adsurl = {https://ui.adsabs.harvard.edu/abs/2015arXiv151006262M}
}

@ARTICLE{Anderson08,
       author = {{Anderson}, Jay and {Sarajedini}, Ata and {Bedin}, Luigi R. and {King}, Ivan R. and {Piotto}, Giampaolo and {Reid}, I. Neill and {Siegel}, Michael and {Majewski}, Steven R. and {Paust}, Nathaniel E.~Q. and {Aparicio}, Antonio and {Milone}, Antonino P. and {Chaboyer}, Brian and {Rosenberg}, Alfred},
        title = "{The Acs Survey of Globular Clusters. V. Generating a Comprehensive Star Catalog for each Cluster}",
      journal = {\aj},
         year = 2008,
        month = jun,
       volume = {135},
       number = {6},
        pages = {2055-2073},
          doi = {10.1088/0004-6256/135/6/2055},
archivePrefix = {arXiv},
       eprint = {0804.2025},
 primaryClass = {astro-ph},
       adsurl = {https://ui.adsabs.harvard.edu/abs/2008AJ....135.2055A}
}

@book{python3,
 author = {Van Rossum, Guido and Drake, Fred L.},
 title = {Python 3 Reference Manual},
 year = {2009},
 isbn = {1441412697},
 publisher = {CreateSpace},
 address = {Scotts Valley, CA}
}

@Article{         numpy,
 title         = {Array programming with {NumPy}},
 author        = {Charles R. Harris and K. Jarrod Millman and St{\'{e}}fan J.
                 van der Walt and Ralf Gommers and Pauli Virtanen and David
                 Cournapeau and Eric Wieser and Julian Taylor and Sebastian
                 Berg and Nathaniel J. Smith and Robert Kern and Matti Picus
                 and Stephan Hoyer and Marten H. van Kerkwijk and Matthew
                 Brett and Allan Haldane and Jaime Fern{\'{a}}ndez del
                 R{\'{i}}o and Mark Wiebe and Pearu Peterson and Pierre
                 G{\'{e}}rard-Marchant and Kevin Sheppard and Tyler Reddy and
                 Warren Weckesser and Hameer Abbasi and Christoph Gohlke and
                 Travis E. Oliphant},
 year          = {2020},
 month         = sep,
 journal       = {Nature},
 volume        = {585},
 number        = {7825},
 pages         = {357--362},
 doi           = {10.1038/s41586-020-2649-2},
 publisher     = {Springer Science and Business Media {LLC}},
 url           = {https://doi.org/10.1038/s41586-020-2649-2}
}

@Article{matplotlib,
  Author    = {Hunter, J. D.},
  Title     = {Matplotlib: A 2D graphics environment},
  Journal   = {Computing in Science \& Engineering},
  Volume    = {9},
  Number    = {3},
  Pages     = {90--95},
  publisher = {IEEE COMPUTER SOC},
  doi       = {10.1109/MCSE.2007.55},
  year      = 2007
}

@software{MESA_11701,
  author       = {Paxton, Bill},
  title        = {{Modules for Experiments in Stellar Astrophysics 
                   (MESA)}},
  month        = may,
  year         = 2019,
  publisher    = {Zenodo},
  version      = {r11701},
  doi          = {10.5281/zenodo.2665077},
  url          = {https://doi.org/10.5281/zenodo.2665077}
}

@sftware{MIST_11701,
    author = {{Dotter}, A and {Bauer}, E},
    title = {MESA for MIST},
    month = April,
    year = 2025,
    publisher = {Zenodo},
    version = {r11701-modified},
    doi       = {10.5281/zenodo.15213405},
    url       = {https://doi.org/10.5281/zenodo.15213405}
}

@software{MESA_work,
    author = {{Dotter}, A and {Bauer}, E},
    title = {MESA Work Directories},
    month = April,
    year = 2025,
    publisher = {Zenodo},
    version = 1,
    doi = {10.5281/zenodo.15232686},
    url = {https://doi.org/10.5281/zenodo.15232686}
}

@ARTICLE{SalarisWeiss1998,
       author = {{Salaris}, M. and {Weiss}, A.},
        title = "{Metal-rich globular clusters in the galactic disk: new age determinations and the relation to halo clusters}",
      journal = {\aap},
         year = 1998,
        month = jul,
       volume = {335},
        pages = {943-953},
archivePrefix = {arXiv},
       eprint = {astro-ph/9802075},
 primaryClass = {astro-ph},
       adsurl = {https://ui.adsabs.harvard.edu/abs/1998A&A...335..943S}
}

@ARTICLE{Basu2004,
       author = {{Basu}, Sarbani and {Mazumdar}, Anwesh and {Antia}, H.~M. and
         {Demarque}, Pierre},
        title = "{Asteroseismic determination of helium abundance in stellar envelopes}",
      journal = {\mnras},
         year = 2004,
        month = may,
       volume = {350},
       number = {1},
        pages = {277-286},
          doi = {10.1111/j.1365-2966.2004.07644.x},
archivePrefix = {arXiv},
       eprint = {astro-ph/0402360},
 primaryClass = {astro-ph},
       adsurl = {https://ui.adsabs.harvard.edu/abs/2004MNRAS.350..277B}
}

@ARTICLE{Nataf2013,
       author = {{Nataf}, David M. and {Gould}, Andrew P. and {Pinsonneault}, Marc H. and
         {Udalski}, Andrzej},
        title = "{Red Giant Branch Bump Brightness and Number Counts in 72 Galactic Globular Clusters Observed with the Hubble Space Telescope}",
      journal = {\apj},
         year = 2013,
        month = apr,
       volume = {766},
       number = {2},
          eid = {77},
        pages = {77},
          doi = {10.1088/0004-637X/766/2/77},
archivePrefix = {arXiv},
       eprint = {1109.2118},
 primaryClass = {astro-ph.GA},
       adsurl = {https://ui.adsabs.harvard.edu/abs/2013ApJ...766...77N}
}

@ARTICLE{Kurucz11,
   author = {{Kurucz}, R.~L.},
    title = "{Including all the lines}",
  journal = {Canadian Journal of Physics},
     year = 2011,
    month = apr,
   volume = 89,
    pages = {417-428},
      doi = {10.1139/p10-104},
   adsurl = {http://adsabs.harvard.edu/abs/2011CaJPh..89..417K}
}

@ARTICLE{Bohlin16,
   author = {{Bohlin}, R.~C.},
    title = "{Perfecting the Photometric Calibration of the ACS CCD Cameras}",
  journal = {\aj},
archivePrefix = "arXiv",
   eprint = {1606.01838},
 primaryClass = "astro-ph.IM",
     year = 2016,
    month = sep,
   volume = 152,
      eid = {60},
    pages = {60},
      doi = {10.3847/0004-6256/152/3/60},
   adsurl = {http://adsabs.harvard.edu/abs/2016AJ....152...60B}
}

@ARTICLE{Paxton18,
       author = {{Paxton}, Bill and {Schwab}, Josiah and {Bauer}, Evan B. and {Bildsten}, Lars and {Blinnikov}, Sergei and {Duffell}, Paul and {Farmer}, R. and {Goldberg}, Jared A. and {Marchant}, Pablo and {Sorokina}, Elena and {Thoul}, Anne and {Townsend}, Richard H.~D. and {Timmes}, F.~X.},
        title = "{Modules for Experiments in Stellar Astrophysics (MESA): Convective Boundaries, Element Diffusion, and Massive Star Explosions}",
      journal = {\apjs},
         year = 2018,
        month = feb,
       volume = {234},
       number = {2},
          eid = {34},
        pages = {34},
          doi = {10.3847/1538-4365/aaa5a8},
archivePrefix = {arXiv},
       eprint = {1710.08424},
 primaryClass = {astro-ph.SR},
       adsurl = {https://ui.adsabs.harvard.edu/abs/2018ApJS..234...34P}
}

@ARTICLE{Paxton13,
   author = {{Paxton}, B. and {Cantiello}, M. and {Arras}, P. and others},
    title = "{Modules for Experiments in Stellar Astrophysics (MESA): Planets, Oscillations, Rotation, and Massive Stars}",
  journal = {\apjs},
archivePrefix = "arXiv",
   eprint = {1301.0319},
 primaryClass = "astro-ph.SR",
     year = 2013,
    month = sep,
   volume = 208,
      eid = {4},
    pages = {4},
      doi = {10.1088/0067-0049/208/1/4},
   adsurl = {http://adsabs.harvard.edu/abs/2013ApJS..208....4P}
}

@ARTICLE{Paxton15,
   author = {{Paxton}, B. and {Marchant}, P. and {Schwab}, J. and others},
    title = "{Modules for Experiments in Stellar Astrophysics (MESA): Binaries, Pulsations, and Explosions}",
  journal = {\apjs},
archivePrefix = "arXiv",
   eprint = {1506.03146},
 primaryClass = "astro-ph.SR",
     year = 2015,
    month = sep,
   volume = 220,
      eid = {15},
    pages = {15},
      doi = {10.1088/0067-0049/220/1/15},
   adsurl = {http://adsabs.harvard.edu/abs/2015ApJS..220...15P}
}

@ARTICLE{Paxton19,
       author = {{Paxton}, Bill and {Smolec}, R. and {Schwab}, Josiah and {Gautschy}, A. and {Bildsten}, Lars and {Cantiello}, Matteo and {Dotter}, Aaron and {Farmer}, R. and {Goldberg}, Jared A. and {Jermyn}, Adam S. and {Kanbur}, S.~M. and {Marchant}, Pablo and {Thoul}, Anne and {Townsend}, Richard H.~D. and {Wolf}, William M. and {Zhang}, Michael and {Timmes}, F.~X.},
        title = "{Modules for Experiments in Stellar Astrophysics (MESA): Pulsating Variable Stars, Rotation, Convective Boundaries, and Energy Conservation}",
      journal = {\apjs},
         year = 2019,
        month = jul,
       volume = {243},
       number = {1},
          eid = {10},
        pages = {10},
          doi = {10.3847/1538-4365/ab2241},
archivePrefix = {arXiv},
       eprint = {1903.01426},
 primaryClass = {astro-ph.SR},
       adsurl = {https://ui.adsabs.harvard.edu/abs/2019ApJS..243...10P}
}

@ARTICLE{Dotter16,
   author = {{Dotter}, A.},
    title = "{MESA Isochrones and Stellar Tracks (MIST) 0: Methods for the Construction of Stellar Isochrones}",
  journal = {\apjs},
archivePrefix = "arXiv",
   eprint = {1601.05144},
 primaryClass = "astro-ph.SR",
     year = 2016,
    month = jan,
   volume = 222,
      eid = {8},
    pages = {8},
      doi = {10.3847/0067-0049/222/1/8},
   adsurl = {http://adsabs.harvard.edu/abs/2016ApJS..222....8D}
}

@ARTICLE{Tolstoy09,
   author = {{Tolstoy}, E. and {Hill}, V. and {Tosi}, M.},
    title = "{Star-Formation Histories, Abundances, and Kinematics of Dwarf Galaxies in the Local Group}",
  journal = {\araa},
archivePrefix = "arXiv",
   eprint = {0904.4505},
 primaryClass = "astro-ph.CO",
     year = 2009,
    month = sep,
   volume = 47,
    pages = {371-425},
      doi = {10.1146/annurev-astro-082708-101650},
   adsurl = {http://adsabs.harvard.edu/abs/2009ARA%26A..47..371T}
}

@ARTICLE{Choi16,
   author = {{Choi}, J. and {Dotter}, A. and {Conroy}, C. and others},
    title = "{Mesa Isochrones and Stellar Tracks (MIST). I. Solar-scaled Models}",
  journal = {\apj},
archivePrefix = "arXiv",
   eprint = {1604.08592},
 primaryClass = "astro-ph.SR",
     year = 2016,
    month = jun,
   volume = 823,
      eid = {102},
    pages = {102},
      doi = {10.3847/0004-637X/823/2/102},
   adsurl = {http://adsabs.harvard.edu/abs/2016ApJ...823..102C}
}

@ARTICLE{Planck15,
   author = {{Planck Collaboration} and {Ade}, P.~A.~R. and {Aghanim}, N. and 
	{Arnaud}, M. and {Ashdown}, M. and {Aumont}, J. and {Baccigalupi}, C. and 
	{Banday}, A.~J. and {Barreiro}, R.~B. and {Bartlett}, J.~G. and et al.},
    title = "{Planck 2015 results. XIII. Cosmological parameters}",
  journal = {ArXiv:1502.01589},
archivePrefix = "arXiv",
   eprint = {1502.01589},
     year = 2015,
    month = feb,
   adsurl = {http://adsabs.harvard.edu/abs/2015arXiv150201589P}
}

@ARTICLE{VandenBerg12,
   author = {{VandenBerg}, D.~A. and {Bergbusch}, P.~A. and {Dotter}, A. and 
	{Ferguson}, J.~W. and {Michaud}, G. and {Richer}, J. and {Proffitt}, C.~R.
	},
    title = "{Models for Metal-poor Stars with Enhanced Abundances of C, N, O, Ne, Na, Mg, Si, S, Ca, and Ti, in Turn, at Constant Helium and Iron Abundances}",
  journal = {\apj},
archivePrefix = "arXiv",
   eprint = {1206.1820},
 primaryClass = "astro-ph.SR",
     year = 2012,
    month = aug,
   volume = 755,
      eid = {15},
    pages = {15},
      doi = {10.1088/0004-637X/755/1/15},
   adsurl = {http://adsabs.harvard.edu/abs/2012ApJ...755...15V}
}

@ARTICLE{Salaris93,
   author = {{Salaris}, M. and {Chieffi}, A. and {Straniero}, O.},
    title = "{The alpha-enhanced isochrones and their impact on the FITS to the Galactic globular cluster system}",
  journal = {\apj},
     year = 1993,
    month = sep,
   volume = 414,
    pages = {580-600},
      doi = {10.1086/173105},
   adsurl = {http://adsabs.harvard.edu/abs/1993ApJ...414..580S}
}

@ARTICLE{Pritzl05,
   author = {{Pritzl}, B.~J. and {Venn}, K.~A. and {Irwin}, M.},
    title = "{A Comparison of Elemental Abundance Ratios in Globular Clusters, Field Stars, and Dwarf Spheroidal Galaxies}",
  journal = {\aj},
   eprint = {arXiv:astro-ph/0506238},
     year = 2005,
    month = nov,
   volume = 130,
    pages = {2140-2165},
      doi = {10.1086/432911},
   adsurl = {http://adsabs.harvard.edu/abs/2005AJ....130.2140P}
}

@ARTICLE{Conroy13b,
   author = {{Conroy}, C.},
    title = "{Modeling the Panchromatic Spectral Energy Distributions of Galaxies}",
  journal = {\araa},
archivePrefix = "arXiv",
   eprint = {1301.7095},
 primaryClass = "astro-ph.CO",
     year = 2013,
    month = aug,
   volume = 51,
    pages = {393-455},
      doi = {10.1146/annurev-astro-082812-141017},
   adsurl = {http://adsabs.harvard.edu/abs/2013ARA%26A..51..393C}
}

@ARTICLE{Caffau11,
   author = {{Caffau}, E. and {Ludwig}, H.-G. and {Steffen}, M. and {Freytag}, B. and 
	{Bonifacio}, P.},
    title = "{Solar Chemical Abundances Determined with a CO5BOLD 3D Model Atmosphere}",
  journal = {\solphys},
archivePrefix = "arXiv",
   eprint = {1003.1190},
 primaryClass = "astro-ph.SR",
     year = 2011,
    month = feb,
   volume = 268,
    pages = {255-269},
      doi = {10.1007/s11207-010-9541-4},
   adsurl = {http://adsabs.harvard.edu/abs/2011SoPh..268..255C}
}

@ARTICLE{Paxton11,
   author = {{Paxton}, B. and {Bildsten}, L. and {Dotter}, A. and {Herwig}, F. and 
	{Lesaffre}, P. and {Timmes}, F.},
    title = "{Modules for Experiments in Stellar Astrophysics (MESA)}",
  journal = {\apjs},
archivePrefix = "arXiv",
   eprint = {1009.1622},
 primaryClass = "astro-ph.SR",
     year = 2011,
    month = jan,
   volume = 192,
    pages = {3},
      doi = {10.1088/0067-0049/192/1/3},
   adsurl = {http://adsabs.harvard.edu/abs/2011ApJS..192....3P}
}

@ARTICLE{Grevesse98,
   author = {{Grevesse}, N. and {Sauval}, A.~J.},
    title = "{Standard Solar Composition}",
  journal = {\ssr},
     year = 1998,
    month = may,
   volume = 85,
    pages = {161-174},
      doi = {10.1023/A:1005161325181},
   adsurl = {http://adsabs.harvard.edu/abs/1998SSRv...85..161G}
}

@ARTICLE{Asplund09,
   author = {{Asplund}, M. and {Grevesse}, N. and {Sauval}, A.~J. and {Scott}, P.
	},
    title = "{The Chemical Composition of the Sun}",
  journal = {\araa},
archivePrefix = "arXiv",
   eprint = {0909.0948},
 primaryClass = "astro-ph.SR",
     year = 2009,
    month = sep,
   volume = 47,
    pages = {481-522},
      doi = {10.1146/annurev.astro.46.060407.145222},
   adsurl = {http://adsabs.harvard.edu/abs/2009ARA%26A..47..481A}
}

@ARTICLE{Chabrier97,
   author = {{Chabrier}, G. and {Baraffe}, I.},
    title = "{Structure and evolution of low-mass stars}",
  journal = {\aap},
   eprint = {arXiv:astro-ph/9704118},
     year = 1997,
    month = nov,
   volume = 327,
    pages = {1039-1053},
   adsurl = {http://adsabs.harvard.edu/abs/1997A%26A...327.1039C}
}

@ARTICLE{Kurucz1970,
   author = {{Kurucz}, R.~L.},
    title = "{Atlas: a Computer Program for Calculating Model Stellar Atmospheres}",
  journal = {SAO Special Report},
     year = 1970,
    month = feb,
   volume = 309,
   adsurl = {http://adsabs.harvard.edu/abs/1970SAOSR.309.....K}
}
\bibliographystyle{aasjournal}

\end{document}